\RequirePackage{fix-cm}
\documentclass[twocolumn]{svjour3}          
\smartqed  
\usepackage{amsfonts}
\usepackage{times,amsmath,epsfig}
\usepackage{epsfig,tabularx,amssymb,amsmath,subfigure,multirow, algorithm, algorithmic,graphicx}
\usepackage{pifont}
\usepackage{cases}
\usepackage{xspace}
\usepackage{color}
\usepackage{hhline}
\usepackage{paralist}
\usepackage{cite}
\usepackage{url}
\makeatletter
\g@addto@macro{\UrlBreaks}{\UrlOrds}
\makeatother

\renewcommand{\algorithmicrequire}{\textbf{Input:}}
\renewcommand{\algorithmicensure}{\textbf{Output:}}

\newtheorem{updaterule}{Update Rule}

\journalname{The VLDB Journal}
\begin{document}

%
\title{Geo-Social Group Queries with Minimum Acquaintance Constraints}

\author{Qijun Zhu \and Haibo Hu \and Cheng Xu \and Jianliang Xu \and Wang-Chien Lee}

\institute{Qijun Zhu \and Cheng Xu \and  Jianliang Xu \at
              Department of Computer Science,\\
              Hong Kong Baptist University,\\
              Kowloon Tong, Hong Kong\\
              \email{\{qjzhu,chengxu,xujl\}@comp.hkbu.edu.hk}
           \and
           Haibo Hu \at
              Department of Electronic and Information Engineering,\\
              Hong Kong Polytechnic University,\\
              Hung Hom, Hong Kong\\
              \email{haibo.hu@polyu.edu.hk}
            \and
           Wang-Chien Lee \at
              Department of Computer Science and Engineering,\\
               Pennsylvania State University, \\
              University Park, USA\\
              \email{wlee@cse.psu.edu}
}
\date{Received: date / Accepted: date}

\maketitle

\begin{abstract}
The prosperity of location-based social networking has paved the way for new applications of group-based activity planning and
marketing. While such applications heavily rely on geo-social group queries (GSGQs), existing studies fail to produce a cohesive group in terms of user acquaintance. In this paper, we propose a new family of GSGQs with minimum acquaintance constraints. They are more
appealing to users as they guarantee a worst-case acquaintance level in the result group.
For efficient processing of GSGQs on large location-based social networks, we devise two
social-aware spatial index structures, namely SaR-tree and SaR*-tree. The
latter improves on the former by considering both
spatial and social distances when clustering objects. Based on SaR-tree and SaR*-tree,
novel algorithms are developed to process various GSGQs.
Extensive experiments on real datasets Gowalla and Twitter show that our
proposed methods substantially outperform the baseline algorithms
under various system settings.
\end{abstract}

\section{Introduction} \label{sec:intro}
With the ever-growing popularity of smartphone devices, the past few
years have witnessed a massive boom in location-based social
networking services (LBSN)~\cite{Zhang15,Hao15,Schlegel15,Khalid15} like Foursquare, Yelp, Google+, and Facebook Places. In
all these applications, mobile users are allowed to share their check-in
locations (e.g., restaurants, theaters) with friends. Such location
information, bridging the gap between the physical world and the
virtual world of social networks, presents to users new applications of group-based activity planning and marketing~\cite{yang12,yafei15,yafei17}. In a typical use case, Facebook now offers users to create or participate a local group event, such as a lunch gathering or a tennis match. With location information, Facebook can proactively recommend users nearby and invite them to this event. Third-party apps can also make use of such information. For example Zimride, on Facebook suggests ridesharing among a group of users with similar commutes.  These location-based social networking applications are essentially \emph{geo-social group queries} with both spatial and social constraints.

While research attention has recently been drawn to geo-social group
queries (e.g., \cite{liu,yang12}), existing works only impose
some \emph{loose} social constraint on the query. For example in \cite{liu}, the circle-of-friend query targets at finding a set of
$k$ users such that the maximal weighted spatial and social distance
among the users is minimized. Since social distance is only one of the two factors, users in the result group could
have very distant or diverse social relations. In an extreme case, no users in the result group are familiar with one another but they are so spatially close that the overall intra-group distance is minimum. As an
improvement, the socio-spatial group query proposed in \cite{yang12}
aims to find $k$ spatially close users among which the
\emph{average} number of unfamiliar users does not exceed a
threshold $p$. While the use of threshold $p$ effectively reduces
the occurrence of unfamiliar users in a result group, there is
no guarantee on the \emph{minimum} number of users a group member is familiar with. In the worst case, as shown in our experiments in
Section~\ref{sec:overall}, some user may be unfamiliar with all other users in the group. Moreover, both queries require tailor-made user inputs --- \cite{liu} imposes weights on social and spatial distances, and \cite{yang12} needs to set a unified threshold $p$ for all users in the group even though different users
may have varied tolerance of unfamiliar users surrounded. Finally, these works mainly focused on
in-memory processing (e.g., improving the user scanning order and
filtering the candidate combinations), and cannot be adapted to external-memory indexes. Therefore, they cannot work for large-scale and real-world LBSNs.

\begin{figure}
\centering
\includegraphics[width=0.78\linewidth]{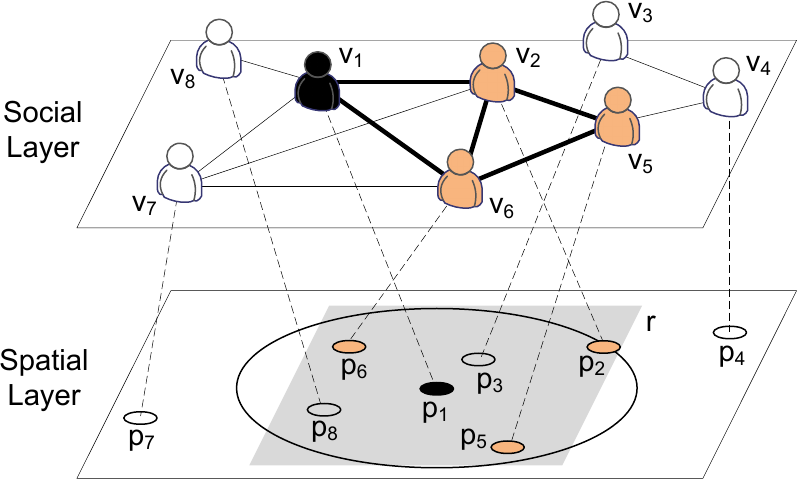}
\caption{An example of GSGQ$<$$v_1,3NN,2$$>$. Lines between the users represent
acquaintance relations and the points on the spatial layer denote
the positions of the users. } \label{fig:gsgq} 
\end{figure}

In this paper, we propose a new family of Geo-Social Group Queries
with constraint on \emph{minimum} acquaintance, hereafter called GSGQs for brevity. A GSGQ query takes three arguments: $(q,
\Lambda, c)$, where $q$ is the query issuer, $\Lambda$ is the
spatial constraint, and $c$ is the acquaintance constraint. The
acquaintance constraint $c$ imposes a minimum degree on the
familiarity of group members (which may include $q$), i.e., every user in
the group should be familiar with at least $c$ other users.
The minimum degree constraint is an important measure of group
cohesiveness in social science research~\cite{seidman}. Known as
$c$-core, it has been widely investigated in the research of graph
problems~\cite{balas, cheng, moser} and accepted as an important
constraint in practical applications~\cite{sozio}.
The spatial constraint $\Lambda$ can be a range constraint, a
$k$-nearest-neighbor ($k$NN) constraint or a
relaxed $k$-nearest-neighbor (r$k$NN) constraint, 
where $k$NN (resp. r$k$NN) means the result group, among all valid groups of exactly (resp. no fewer than) $k$ users that satisfy the minimum acquaintance constraint, has the minimum spatial distance to the query issuer.

Fig.~\ref{fig:gsgq} illustrates an example of GSGQ, where the social
network is split into a social layer and a spatial layer for clarify
of presentation. Suppose user $v_1$ wants to arrange a friend
gathering of some friends nearby. To have a friendly atmosphere in
the gathering, she hopes anyone in the group should be
familiar with at least two other users. Thus, she issues a GSGQ
$=(q,\Lambda,c)$ with $q$ set as $v_1$, $\Lambda$ being $3$NN, and
$c=2$. With the objective of minimizing the spatial distance between
$q$ and the farthest user in the group, the result group she will obtain
is $W = \{v_2, v_5, v_6\}$.
Alternatively, to find an acquainted
group of friends within a fixed range, she may issue a GSGQ $=
(q,\Lambda,c)$ with $q$ set as $v_1$, $\Lambda$ being $r$ (shaded area in Fig.~\ref{fig:gsgq}), and
$c=2$. In this example, she will also obtain $W = \{v_2, v_5,
v_6\}$.

We argue that, compared to the geo-social group queries studied in
prior work~\cite{liu,yang12}, our GSGQs, with the adoption of a
minimum acquaintance constraint, are more appealing to produce a
cohesive group that guarantees the worst-case acquaintance level.
Nonetheless, these GSGQs are much more complex to process than conventional spatial
queries. Particularly, when the spatial constraint is strict $k$NN, we
prove that GSGQs are NP-hard. Due to the additional social
constraint, traditional spatial query processing techniques
\cite{finkel, guttm, beckmann, papad} cannot be directly applied to
GSGQs. Moreover, these queries are intrinsically harder than other variants of spatial queries, such as
spatial-keyword queries \cite{feli, zhang, wu} and collective
spatial keyword queries \cite{cao}, which only introduce
\emph{independent} attributes (e.g., text descriptions) of the
objects but not binary relations among them.

On the other hand, most previous works on group queries in social
networks use sequential scan in query processing. That is, they
enumerate every possible combination of a user group and optimize
the processing through some pruning heuristics. Although
\cite{yang12} proposed an SR-tree to cluster the users of each leaf
node, this index achieves more significant reduction on computation than on disk accesses since it separates spatial and social constraints in the clustering process. Thus, when geo-social queries such as GSGQs are processed, still many disk pages are accessed to fetch the users that satisfy both spatial and social constraints. Moreover, its filtering
techniques only work for average-degree social constraints, and are
not suitable for GSGQs with minimum-degree social constraints.
In this paper, we propose two novel
social-aware spatial indexing structures, namely, SaR-tree and
SaR*-tree, for efficient processing of general GSGQ queries on
external storage. The main idea is to project the social relations
of an LBSN on the spatial layer and then index both social and
spatial relations in a uniform tree structure to facilitate GSGQ
processing. Furthermore, we optimize the in-memory processing of
GSGQs with a strict $k$NN constraint by devising powerful pruning
strategies. To sum up, the main contributions of this paper are as
follows:

\begin{itemize}
 \item We propose a new family of geo-social group queries with
minimum acquaintance constraint (GSGQs), which guarantees the
worst-case acquaintance level. We prove that the GSGQs with a
strict $k$NN spatial constraint are NP-hard.
 \item We design new social-aware index structures, namely SaR-tree
and SaR*-tree, for GSGQs. To optimize the I/O access and processing
cost, a novel clustering technique that considers both spatial and
social factors is proposed in the SaR*-tree. The update procedures of
both indexes are also presented.
 \item Based on the SaR-tree and SaR*-tree, efficient algorithms are
 developed to process various GSGQs.
Moreover, in-memory optimizations are proposed
for GSGQs with a strict $k$NN constraint.
 \item We conduct extensive experiments to demonstrate the performance
of our proposed indexes and algorithms.
\end{itemize}

The rest of this paper is organized as follows.
Section~\ref{related} reviews the related works.
Section~\ref{problem} introduces some core concepts in the social
constraint and formalizes the problems of GSGQs.
Section~\ref{srtree} presents the designs of basic SaR-tree and
optimized SaR*-tree. Section~\ref{app} details the processing
methods for various GSGQs based on SaR-trees. Section~\ref{update}
describes the update algorithms of SaR-trees.
Section~\ref{experiment} evaluates the performance of our proposals.
Finally, Section~\ref{conclude} concludes the paper and discusses
future directions.

\section{Related Works} \label{related}

\subsection{Spatial Query Processing}
\begin{figure}
\centering
\includegraphics[width=0.9\linewidth]{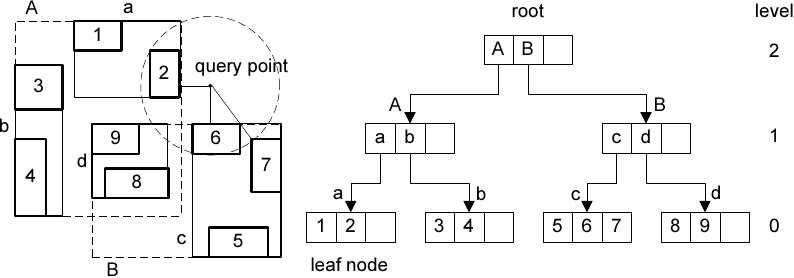}
\caption{An example of R-tree. }  \label{fig:3NNQuery} 
\end{figure}

Many spatial databases use R-tree or its extensions~\cite{guttm,
beckmann} as an access method to disk storage for spatial queries (e.g.,
range, $k$NN, and spatial join queries). Fig.~\ref{fig:3NNQuery}
shows nine objects in a two-dimensional space and how they are
aggregated into Minimum Bounding Rectangles (MBRs) recursively to
build up the corresponding R-tree. An R-tree node is composed of a
number of entries, each covering a set of objects and using an MBR
to bound them. A query is processed by traversing the R-tree from
the root node all the way down to leaf nodes for qualified objects.
During this process, a priority queue $H$ can be used to maintain
the entries to be explored. A generic query evaluation procedure for
a query $Q$ can be summarized as follows: (1) push the entries of the root node
into $H$; (2) pop up the top entry $e$ from $H$; (3) if $e$ is a
leaf entry, check if the corresponding object is a result object;
otherwise push all qualified child entries of $e$ into $H$; (4)
repeat (2) and (3) until $H$ is empty or a termination condition of
$Q$ is satisfied. The construction of R-trees can be either incremental~\cite{guttm,
beckmann} or bulk-loaded.


Some variants of spatial queries have been studied with the
consideration of certain grouping semantics. The \emph{group
nearest-neighbor} query \cite{papad} extends the concept of the
nearest neighbor query by considering a group of query points. It
targets at finding a set of data points with the smallest sum of
distances to all the query points. Based on R-tree, \cite{papad}
proposes various pruning heuristics to efficiently process group
nearest-neighbor queries. The \emph{spatial-keyword} query is
another well-known extension of spatial queries that exploits both
locations and textual descriptions of the objects. Most solutions
for this query, e.g., BR*-tree \cite{zhang},  $IR^2$-tree
\cite{feli}, and IR-tree \cite{wu}, rely on combining the inverted
index, which was designed for keyword search, with a conventional
R-tree. The \emph{collective spatial keyword} query \cite{cao}
further considers the problem of retrieving a group of spatial web
objects such that the group's keywords cover the query's keywords
and the objects, with the shortest inter-object distances, are nearest
to the query location. Based on IR-tree, \cite{cao} proposes dynamic
programming algorithms for exact query processing and greedy
algorithms for approximate query processing. It is noteworthy that,
while these works deal with some grouping semantics, they do not
consider acquaintance relations in social networks.

\subsection{Social Network Analysis and Query Processing}
There have been a lot of works on community discovery in social
networks. There is a comprehensive survey on community finding in graphs~\cite{fortunato}. A typical approach is to optimize the modularity
measure~\cite{girvan}. Since communities are usually cohesive
subgraphs formed by the users with the acquaintance relationship,
some graph structures such as clique \cite{harary},
$k$-core~\cite{seidman}, and $k$-plex~\cite{balas, mcclosky, moser}
have been well studied under this topic. However, most of these
works only provide theoretical solutions with asymptotic complexity,
with a few exceptions such as the external-memory top-down algorithm
for core decomposition \cite{cheng}.

As for query processing in social networks, \cite{faloutsos}
addresses the problem of finding a subgraph that connects a set of
query nodes in a graph. \cite{sozio} studies a query-dependent
variant of the community discovery problem, which finds a dense
subgraph that contains the query nodes. 
Based on a measure of graph density, an optimal greedy algorithm is
proposed. The authors of \cite{sozio} also prove that finding
communities of size no larger than a specified upper bound is
NP-hard. Besides, \cite{yang} proposes a social-temporal group query
with acquaintance constraint in social networks. The aim is to find
the activity time and attendees with the minimum total social
distance to the initiator. As this problem is NP-hard,
heuristic-based algorithms have been proposed to reduce the run-time
complexity. However, all these works do not consider the spatial
dimension of the users and thus cannot be applied to location-based
social networks.

\subsection{Geo-Social Query Processing}
Efficient processing of queries that consider both spatial and
social relations is essential for LBSNs. \cite{doyt} directly
combines spatial and social networks and proposes graph-based query
processing techniques. \cite{liu} proposes a
circle-of-friend query to find minimal-diameter social groups. By
transforming the relations in social networks into social distances
among users, an integrated distance combining both spatial and
social distances is proposed. \cite{yang12} considers a special
socio-spatial group query with the requirement of minimizing the
total spatial distance. Accordingly, in-memory pruning and searching schemes are proposed in \cite{yang12}. All these works only impose
a \emph{loose} social constraint in the query. As for the processing
techniques, the methods of these works enumerate all possible
combinations guided by some searching and pruning schemes. Although
a tree structure named SR-tree is introduced in \cite{yang12}, it is
mainly used to reduce the enumeration of states during the in-memory processing.
With that said, external-memory indexes tailored for geo-social query
processing in large-scale LBSNs are still lacking. More recently, \cite{Armenatzoglou} proposes a general framework for geo-social query processing, which separates the social, geographical and query processing modules and thus enables flexible data management. Since its pruning power comes separately from the social and spatial index, it cannot further optimize the processing of GSGQ with access methods that integrate both spatial and social information. \cite{yafei15} studies a geo-social query that retrieves a group of socially connected users whose familiar regions collectively cover a set of query points. \cite{Zhang13} proposes a geo-social location recommendation system based on personalized social and geographical influence modeling. Similarly, \cite{Shi14} proposes to cluster and categorize locations based on social and spatial density obtained from geo-social networks.

\section{Preliminaries and Problem Statement} \label{problem}
Aiming to find a cohesive group of acquaintances, GSGQs use
$c$-core~\cite{seidman} as the basis of social constraint to
restrict the result group. In this section, we first introduce the
definition and the properties of $c$-core, based on which the GSGQ
problems are then formalized.

\subsection{$C$-Core}
$c$-core is a degree-based relaxation of clique~\cite{seidman}.
Consider an undirected graph $G = (V, E)$, where $V$ is the set of
vertices and $E$ is the set of edges. Given a vertex $v\in V$, we
define the set of neighbors of $v$ as $N_G(v) = \{u \in V~|~uv \in
E\}$ and the degree of $v$ as $deg_G(v) = |N_G(v)|$. Accordingly,
the maximum and minimum degrees of $G$ are represented as $\Delta(G)
= max_{v\in V}{deg_G(v)}$ and $\delta(G) = min_{v\in V}{deg_G(v)}$,
respectively. Let $G[W]$ denote a subgraph induced by $W \subseteq
V$. The following is a generalized definition of a
$c$-core~\cite{seidman}.

\begin{definition}  \label{def:kcore}
\textbf{($c$-core)} A subgraph $G[W]$ is a
\emph{c-core} (or \emph{a core of order c}) if $\delta(G[W]) \geq c$.
\end{definition}

The $c$-core defined in Definition~\ref{def:kcore} is not required
to be maximum and fits for GSGQs in various applications. In the
sequel, the term $c$-core refers to both the set $W$ and the
subgraph $G[W]$. The \emph{core number} of a vertex $v$, denoted by
$c_v$, is the highest order of a core that contains this vertex.

A greedy algorithm can be used for \emph{core decomposition}, i.e.,
finding the core numbers for all vertices in $G$. The basic idea is
to iteratively remove the vertex with the minimum degree in the
remaining subgraph, together with all the edges adjacent to it, and
determine the core number of that vertex accordingly. The most
costly step of this algorithm is sorting the vertices according to
their degrees at each iteration. As shown in \cite{batag}, a
bin-sort can be used with $O(|V|+|E|)$ time complexity. Thus, for a
given $c$, we can find the maximum $c$-core of $G$ in $O(|V|+|E|)$
time.

\subsection{Problem Statement}
Consider an LBSN $G = (V, E)$, where the set of vertices $V$ denotes
the users and the set of edges $E$ denotes the acquaintance
relations\footnote{Such relation can be either a ``friend" relation or a more intimate acquaintance relation, depending on the nature of the group event in a GSGQ service.} among the users in $V$. For any two users $v, u\in V$,
there exists an edge $vu \in E$ if and only if $v$ is acquainted
with $u$. Moreover, for any user $v\in V$, its location $p_v$ is
also stored in $G$. Given two users $v$ and $u$, let $d(v, u)$
denote the spatial distance between $v$ and $u$, and the (largest)
distance from $v$ to a set of users $W$ is
defined by $d_{max}(v, W) = max_{u\in W}{d(v, u)}$.

As formally defined below, a GSGQ finds a group of users that
satisfies the given spatial and social constraints. Without loss of
generality, we assume that the query issuer $q\in V$.
\begin{definition}  \label{def:gsgq}
\textbf{(Geo-Social Group Query with Minimum Acquaintance Constraint
(GSGQ))} Given an LBSN $G=(V, E)$, a GSGQ is represented as
$Q_{gs}=(v, \Lambda, c)$, where $v\in V$ is the query issuer,
$\Lambda$ is a type of spatial query denoting the spatial
constraint, and $c$ is the minimum degree of result group, denoting
the social acquaintance constraint as in~\cite{liu,yang12}. GSGQ finds a maximal user result set $W$
which satisfies $\Lambda$ and the condition
that the induced subgraph $G[W\cup \{v\}]$ is a $c$-core, or formally,
$\delta(G[W\cup \{v\}])~\geq~c$.
\end{definition}

As for the spatial constraint, this paper mainly focuses on three
query types: range (i.e., window) query, relaxed k-nearest-neighbor (r$k$NN) query,
and strict k-nearest-neighbor ($k$NN) query. Accordingly, they
correspond to three types of GSGQs:
\begin{compactitem}
  \item GSGQ with range constraint, denoted as $GSGQ_{range}$. A
$GSGQ_{range}$ is represented as $Q_{gs}=(v, range, c)$, where
$p_v\in range$. It targets at finding the largest $c$-core $W\cup
\{v\}$ located inside $range$, a rectangular spatial window. For example, ``find me the largest user group satisfying $c$-core in 5th Avenue, Manhattan, NYC."
  \item GSGQ with relaxed $k$NN constraint, denoted as $GSGQ$\hfill$_{rkNN}$. A
$GSGQ_{rkNN}$ is represented as $Q_{gs}=(v, rkNN,$ $c)$. It targets at
finding a maximal $c$-core $W\cup \{v\}$ of size no less than $k+1$ with the
minimum $d_{max}(v, W)$. Here ``relaxed" means the size of the result is not strictly $k+1$, and as a general requirement in GSGQ the size should be the largest possible. For example, ``find me the closest (maximal) group of at least 9 users satisfying $c$-core to be eligible for a bulk discount."
  \item GSGQ with strict $k$NN constraint, denoted as $GSGQ_{kNN}$. A
$GSGQ_{kNN}$ is represented as $Q_{gs}=(v,$ $kNN,c)$. It is a strict
form of $GSGQ_{rkNN}$, which requires that the $c$-core $W\cup
\{v\}$ has an exact size of $k+1$. For example, ``find me the closest group of 3 users satisfying $c$-core to play tennis doubles with me."
\end{compactitem}

For these GSGQs, we prove the following theorems on their
complexities.

\begin{theorem}
$GSGQ_{range}$ and $GSGQ_{rkNN}$ can be solved in polynomial time.
\end{theorem}

\begin{proof}
As we will show in the next subsection, processing a $GSGQ_{range}$
can be completed by running core-decomposition once, while
processing a $GSGQ_{rkNN}$ can be completed by running
core-decomposition at most $|V|$ times. Since the time complexity of
core-decomposition is $O(|V|+|E|)$, both of the queries can be
solved in polynomial time.
\end{proof}

\begin{theorem}
$GSGQ_{kNN}$ is NP-hard.
\end{theorem}
\begin{proof}
It has been proved in \cite{balas} that, given a graph $G$ and
positive integers $\bar{c}$ and $k$, determining whether there
exists a $\bar{c}$-plex of size $k+1$, i.e., a set $W$ such that
$\delta(G[W]) \geq |W| - \bar{c}$ and $|W| = k+1$, is NP-complete.
Since a $c$-core of size $k+1$ is equivalent to a $(k+1-c)$-plex, we
can find a $(k+1-c)$-plex of size $k+1$ by iteratively applying
$GSGQ_{kNN}$ for each user $v$ in $G$. If a $c$-core of size $k+1$
is found for a user $v$, then a $(k+1-c)$-plex of size $k+1$ exists;
otherwise such a $(k+1-c)$-plex does not exist. In this way, the
$\bar{c}$-plex problem can be polynomially reduced to $GSGQ_{kNN}$.
This proves that $GSGQ_{kNN}$ is NP-hard.
\end{proof}

\subsection{R-tree based Query Processing} \label{pb:rt}
We consider the GSGQ problems for large-scale LBSNs where the users' location and social 
information are stored separately on external disk storage as described in~\cite{Armenatzoglou}. A baseline approach
of processing GSGQs on an R-tree index of user locations is as
follows. For a $GSGQ_{range}$ $Q_{gs}=(v, range, c)$, we first find
all users located inside $range$ via R-tree, then compute the
$c$-core $W'$ of the subgraph formed by these users. If $v$ exists
in $W'$, then $W=W'-\{v\}$ is the final result; otherwise, there is
no result for $Q_{gs}$. Since the user filtering step can be done in
$O(|V|)$ time and the core decomposition step can be done in
$O(|V|+|E|)$ time, the complexity of this method is $O(|V|+|E|)$.

For a $GSGQ_{rkNN}$ $Q_{gs}=(v, rkNN, c)$, according to its definition, we access the users in
ascending order of their spatial distances to $v$. As such, we use a similar procedure to kNN search on R-tree. Specifically, we employ a priority queue
$H$ whose priority score is spatial distance to $v$, and a candidate result set $\widetilde{W}$. At the beginning, $\widetilde{W}$ is
initialized as $\{v\}$ and all the root entries of the R-tree are
put into $H$. Each time the top entry $e$ of $H$ is popped up and
processed. If $e$ is a non-leaf entry, its child entries are
accessed and put into $H$; otherwise, $e$ is a leaf entry, i.e., a
user, so $e$ is added into $\widetilde{W}$. When the size of
$\widetilde{W}$ exceeds $k$, we compute the $c$-core $W'$ of the
subgraph formed by the users in $\widetilde{W}$. If $|W'| \geq k+1$
and $v\in W'$, $W=W'-\{v\}$ is the result; otherwise, the above
procedure is continued until the result is found. Since each round
of $c$-core detection can be done in $O(|V|+|E|)$ time, the
complexity of this method is $O(|V|(|V|+|E|))$.

For a $GSGQ_{kNN}$ $Q_{gs}=(v, kNN, c)$, the processing is similar
to $GSGQ_{rkNN}$. The major difference is how to find the result
from $\widetilde{W}$. Since the query returns exact $k$ users, all
possible user sets of size $k+1$ and containing $v$ are checked to
see if it is a $c$-core. If such a user set $W'$ exists, then
$W=W'-\{v\}$ is the result. There are $C^{|V-1|}_{k}$ possible user
sets to be checked, where $C^{|V-1|}_{k}$ denotes the number of
$k$-combinations from the user set $V-\{v\}$. Thus, the complexity
of this method is $O(C^{|V-1|}_{k}(|V|+|E|))$,

Obviously, these approaches are inefficient for GSGQs with a large $c$ value, because a large $c$ means tighter social constraints and thus result users from farther away. According to a recent study~\cite{Shin16}, the maximum $c$ of a graph where the $c$-core exists obeys a 3-to-1 power law
with respect to the count of triangles in the graph. This implies that the number of users to search and check in these approaches increases exponentially as $c$ increases. On the other hand, intuitively a large $c$ means higher chances to prune the irrelevant users before finding the result users. As will be proved and shown in the rest of this paper, the efficiency can be significantly improved by filtering the irrelevant users and optimizing the processing order.

\section{Social-aware R-trees} \label{srtree}
Since a GSGQ involves both spatial and social constraints, to
expedite its processing, both spatial locations and social relations
of the users should be indexed simultaneously. Unfortunately, R-tree
only indexes spatial locations of the users and is thus inefficient.
In this section, we design novel Social-aware R-trees (SaR-trees) to
form the basis of our query processing solutions. In what follows,
we first introduce the concept of Core Bounding Rectangle (CBR) and
then present the details of SaR-tree, followed by a variant SaR*-tree.

\begin{figure}
\centering
\includegraphics[width=0.6\linewidth]{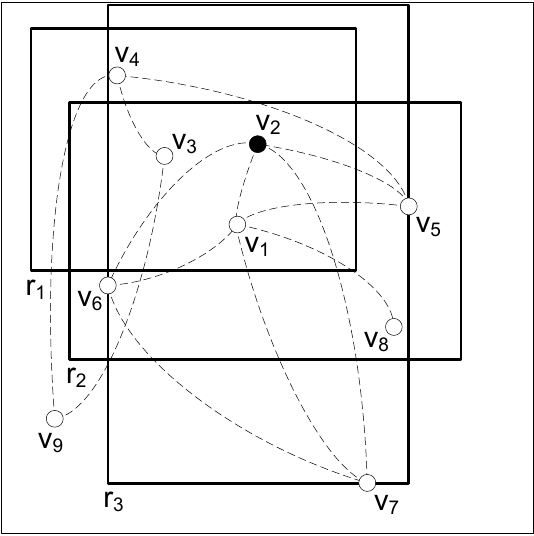}
\caption{An example of CBR. The LBSN is shown on the spatial layer. The points represent
the users as well as their positions, while the dashed lines denote
the acquaintance relations among users. } \label{fig:cbr}
\end{figure}

\subsection{Core Bounding Rectangle (CBR)}
The social constraint of a GSGQ
requires the result group to be a $c$-core. Unfortunately, pure social
measures such as core number and centrality
cannot adequately facilitate GSGQ processing which also features a
spatial constraint.
To devise effective spatial-dependant social measures
to filter users in query processing, in this paper, we develop the
concept of \emph{Core Bounding Rectangle (CBR)} by projecting the
minimum degree constraint on the spatial layer. Simply put, the
CBR of a user $v$ is a rectangle containing $v$, inside which any
user group with $v$ does not satisfy the minimum degree constraint.
In other words, it is a localized
social measure to a user. As a GSGQ mainly requests the nearby
users, the locality of CBR becomes very valuable for processing
GSGQs. The formal description of a CBR of user $v$ for a minimum
degree constraint $c$, denoted by $CBR_{v, c}$, is given in
Definition~\ref{def:cbr}.

\begin{definition}  \label{def:cbr}
\textbf{(Core Bounding Rectangle (CBR))} Consider a user $v \in G$.
Given a minimum degree constraint $c$, $CBR_{v,c}$ is a rectangle
which contains $v$ and inside which any user group with $v$
(excluding the users on the bounding edges) cannot be a $c$-core.
Formally, $CBR_{v,c}$ satisfies $p_v\in CBR_{v,c}$ and $\forall
W=\{v\}\cup \{u|u\in V, p_u\in CBR_{v,c}\}$ $\delta(G[W])<c$.
\end{definition}

An example is shown in Fig.~\ref{fig:cbr}. According to the
acquaintance relations of user $v_2$,
rectangular area $r_1$ is a $CBR$\hfill$_{v_2, 2}$, because any user group
inside $r_1$ that contains $v_2$ cannot be a $2$-core. On the
contrary, $r_2$ is not a $CBR_{v_2, 2}$, because some user groups
inside $r_2$ that contain $v_2$, e.g., $\{v_2, v_1,$\hfill $v_6\}$, are
$2$-cores. Note that $CBR_{v, c}$ is
not unique for a given $v$ and $c$. For example, $r_3$ is another
$CBR_{v_2, 2}$ for user $v_2$. From Definition~\ref{def:cbr}, we
can quickly exclude a user $v$ from the result group by
checking $CBR_{v, c}$ during query processing.
For example, if the query range of a $GSGQ_{range}$ is covered by
$CBR_{v, c}$, then $v$ can be safely pruned from the result.
This property makes CBR a powerful pruning mechanism.

\begin{figure}
  \centering
  \subfigure[Initialization]{
    \label{fig:subfig:initcbr} 
    \includegraphics[width=0.48\linewidth]{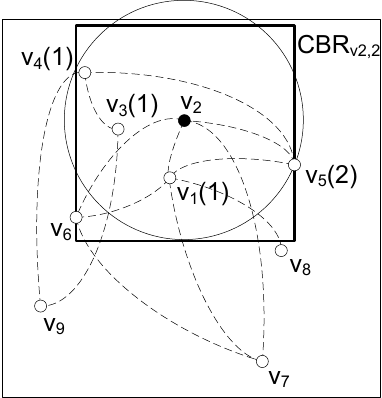}}
  \subfigure[Expansion]{
    \label{fig:subfig:expdcbr} 
    \includegraphics[width=0.48\linewidth]{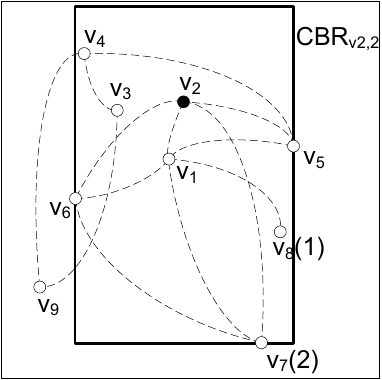}}
  \caption{An
exemplary procedure of computing $CBR_{v_2,2}$ in an LBSN. The
number after a user $v_i$ denotes the core number of $v_2$ in the
subgraph determined by $v_i$. For a), the subgraph is formed by the
users inside $\odot_{v_2, v_i}$; for b), the subgraph is formed by
the users inside $CBR_{v_2,2}$ when moving its bottom edge outward
to go through $v_i$.}
  \label{fig:compcbr} 
\end{figure}

\textbf{Computing CBR of a User}. In an LBSN $G$, given a user $v$
and minimum degree constraint $c$, a simple method to compute
$CBR_{v,c}$ is to search neighboring users in ascending order of distance
until there is a user $u$ such that the core
number of $v$ in the subgraph formed by the users inside $\odot_{v,
u}$ (i.e., the circle centered at $v$ with radius $d(v,u)$) is no less than $c$, i.e., all user groups located within
$\odot_{v, u}$ are not qualified as a $c$-core. $CBR_{v,c}$ can then be
easily derived from $\odot_{v, u}$ as follows. We first compute the bounding box of the circle and move out one bounding edge to go through $u$. Then we check the nodes inside the rectangle but outside the circle. For each of them, we move out one bounding edge to go through $u$ so that the node becomes outside of the new rectangle.
An example is shown in Fig.~\ref{fig:subfig:initcbr}, where a
$CBR_{v_2,2}$ is constructed based on users $v_5$, $v_6$, and $v_8$.
This generated CBR satisfies Definition~\ref{def:cbr} since the
users inside it (i.e., $v_1,v_2,v_3$) cannot form 2-core groups.
However, it is not a maximal one, thus limiting its pruning power in
GSGQ processing. We improve this initial $CBR_{v,c}$ by recursively
expanding it from each bounding edge until no edge can be further
moved outward (see Fig.~\ref{fig:subfig:expdcbr}). Depending on
different initial CBRs and different expanding orders, there could
be a number of maximal CBRs.

Algorithm~\ref{alg:compcbr} details the procedure of computing
$CBR_{v,c}$. In Line 1, we first sort the users of $V$ in ascending
order of their distances to $v$. In Lines 2-5, we find the nearest
user $u$ such that $c_v \geq c$ in the subgraph formed by the users
in $V$ with equal or shorter distances to $v$. In Line 6, we
initialize $CBR_{v,c}$ based on $u$ such that $CBR_{v,c}$ does not
contain any user outside $\odot_{v,u}$. An exemplary way is to
compute the bounding box of $\odot_{v,u}$ first and move one
bounding edge to go through $u$. Then, check the users which are
located inside the rectangle but outside $\odot_{v,u}$. For each of
them, move one bounding edge of the rectangle to go through it so
that the user is not located inside the new rectangle. In this
procedure, a greedy scheme is adopted to always select the bounding
edge which maximizes the area of the rectangle. In Lines 7-10, we
expand $CBR_{v,c}$ by moving each bounding edge $l$ of $CBR_{v,c}$
outward, if $c_v < c$ in the subgraph formed by the users inside
$CBR_{v,c}$ and on $l$. Obviously, the rectangle generated by
Algorithm~\ref{alg:compcbr} is a maximal $CBR_{v,c}$, i.e., it is a
$CBR_{v,c}$ and cannot be fully covered by any other $CBR_{v,c}$.
This property guarantees its pruning power for GSGQ processing, and such maximal $CBR$s will be stored in the social-aware R-trees.
Fig.~\ref{fig:compcbr} provides an exemplary procedure for computing
$CBR_{v_2,2}$ when applying Algorithm~\ref{alg:compcbr} on an LBSN.

\begin{algorithm}
\caption{Computing CBR of a User} \label{alg:compcbr}
\begin{algorithmic}[1]
\footnotesize
\REQUIRE LBSN $G=(V, E)$, user $v$, constraint $c$ \ENSURE $CBR_{v,c}$\\
\emph{CompCBR}($G$, $v$, $c$) \STATE Sort users of $V$ in ascending
order of distances to $v$; \FOR{each user $u$ in $V$} \STATE Compute
$c_v$ in the subgraph formed by the users before (and including)
$u$; \IF{$c_v\geq c$} \STATE Break; \ENDIF \ENDFOR \STATE Build an
initial $CBR_{v,c}$ which goes through $u$ and does not contain any
user outside $\odot_{v, u}$; //$u$ is the user that breaks the above loop \STATE Sort users of $V$ in horizontal
and vertical order, respectively. \WHILE{existing a bounding edge
$l$ of $CBR_{v,c}$ s.t. $c_v < c$ in the subgraph formed by the
users inside $CBR_{v,c}$ and on $l$} \STATE Move $l$ outward to the
next (or previous) user in horizontal (or vertical) order until $c_v
\geq c$ in the subgraph formed by the users inside $CBR_{v,c}$ and
on $l$; \ENDWHILE \STATE return $CBR_{v,c}$;
\end{algorithmic}
\end{algorithm}

To save the computing and storage cost, we only maintain a limited
number of CBRs for user $v$ --- $CBR_{v, 2^0}$, $CBR_{v, 2^1}$,
$\cdots$, $CBR_{v, 2^{\lfloor \log_2{c_v}\rfloor}}$ --- where $c_v$
is the core number of $v$ in $G$.
We choose CBRs with respect to exponential minimum degree
constraints because for a larger $c$, as shown in Section~\ref{experiment}, much
fewer $c$-cores exist and keeping sparse CBRs is sufficient to support effective pruning.


\textbf{Complexity Analysis}. Let $n = |V|$ and $m = |E|$. In
Algorithm~\ref{alg:compcbr}, the sorting step, i.e., Line 1,
requires $O(nlogn)$ time complexity. Since the core number of a user
in graph $G$ can be computed in $O(n + m)$ time, initializing
$CBR_{v,c}$ in Lines 2-6 requires $O((n+m)n)$ time complexity.
Further sorting step in Line 7 requires $O(nlogn)$ time complexity.
During CBR expansion in Lines 8-9, the movement of a bounding edge
requires $O((n+m)n)$ time complexity.
In total, the time complexity of Algorithm~\ref{alg:compcbr} is
$O((n+m)n)$. By applying a binary search to find a proper $u$ in CBR
initialization and a proper user to go through in CBR expansion, the
time complexity can be reduced to $O((n+m)logn)$. Usually, $m>n$ in
an LBSN, so the time complexity of Algorithm~\ref{alg:compcbr} is
$O(mlogn)$.

\subsection{SaR-tree}
\begin{figure}
\centering
\includegraphics[width=\linewidth]{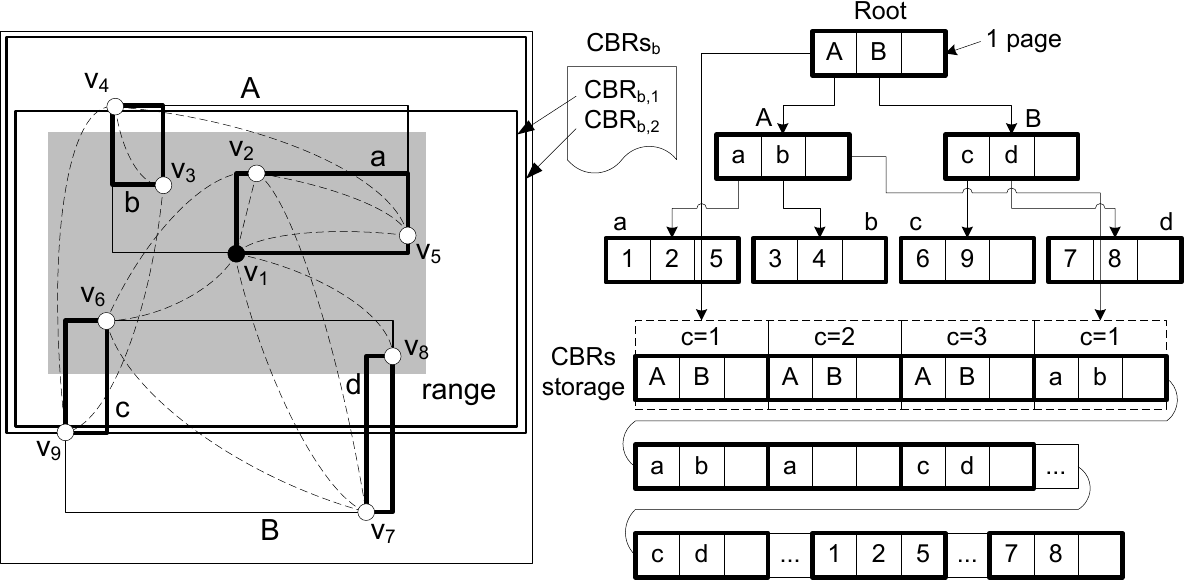}
\caption{SaR-tree. $CBRs_{e}$ denotes the set of CBRs for an entry
$e$.} \label{fig:srtree} 
\end{figure}

We now present the basic SaR-tree. It is a variant of R-tree in
which each entry further maintains some aggregate social-relation
information for the users covered by this entry.
Fig.~\ref{fig:srtree} exemplifies an SaR-tree. Different from a
conventional R-tree, each entry of an SaR-tree refers to two pieces
of information, i.e., a set of CBRs (detailed below) and an MBR, to
describe the group of users it covers. An example of the former,
$CBRs_{b}$ is shown in the figure. It comprises the core number
$c_b$ and two CBRs $\{CBR_{b,1}, CBR_{b,2}\}$ for entry $b$. The
core number of an entry is the maximum core number of the users it
covers, which bounds the number of CBRs of this entry. Considering
that only one CBR of an entry is related to a GSGQ, we optimize the
storage by decoupling CBRs from MBR, as shown in
Fig.~\ref{fig:srtree}. Then, we can directly access the CBR page
with the specified $c$, without losing any pruning power of R-tree.
Perceptually, a CBR in the SaR-tree bounds a group of users from the
social perspective while an MBR bounds the users from the spatial
perspective. As such, SaR-tree gains the power for both social-based
and spatial-based pruning during GSGQ processing, as will be
explained in the next section.

\textbf{CBR of an Entry}. To define the CBRs for each SaR-tree
entry, we extend the concept of CBR defined for each individual user
(in the previous subsection). Let $MBR_e$ and $V_e$ denote the MBR
and the set of users covered by an entry $e$, respectively. A CBR of
$e$ is a rectangle which intersects $MBR_e$ and inside which any
user group containing any user from
$V_e$ cannot satisfy the minimum degree constraint. The formal
definition of a CBR of entry $e$ with respect to a minimum degree
constraint $c$, denoted by $CBR_{e, c}$, is given as follows:

\begin{definition}  \label{def:cbre}
\textbf{(CBR of an Entry)} Consider an entry $e$ with MBR $MBR_e$
and user set $V_e$. Given a minimum degree constraint $c$,
$CBR_{e,c}$ is a rectangle which intersects $MBR_e$ and inside which
any user group containing any user from $V_e$ (not including the users on the bounding edges) cannot
be a $c$-core.
\end{definition}

Note that $CBR_{e,c}$ is
required to intersect $MBR_e$ to guarantee its locality. 
Fig.~\ref{fig:srtree} shows two examples of CBRs for an entry $b$,
where $V_b = \{v_3, v_4\}$. We can see that any user group inside
$CBR_{b,2}$ and containing $v_3$ or
$v_4$ (not including $v_9$ on the bounding edges) cannot be a
$2$-core. Thus, during GSGQ processing, we may safely prune entry
$e$, for example, if the query range of a $GSGQ_{range}$ (with a
minimum degree constraint of 2) is fully covered by $CBR_{b,2}$.
Since $CBR_{e, c}$ is determined by the set of users in $V_e$, we
use $CBR_{V_e, c}$ and $CBR_{e, c}$ interchangeably.



To efficiently generate the CBRs of the entries in
SaR-tree, we adopt a bottom-up approach in our implementation.
Obviously, the CBR of a leaf entry $e$ is just the CBR of the user
it covers. For a non-leaf entry $e$, let $e_1, e_2, \cdots, e_m$ be
the child entries of $e$. Then, the CBR of $e$ can be computed by
recursively applying the following function on $CBR_{e_1},$\hfill $\ldots,
CBR_{e_m}$:
\begin{displaymath}CBR_{\{e_1, \ldots, e_{i+1}\},c}=\left\{
\begin{array}{ll}
    CBR_{\{e_1, \ldots, e_i\},c}, \text{if }~~~~~~~\\ ~~~~~\mbox{$MBR_{e_{i+1}}\cap CBR_{\{e_1, \ldots, e_i\},c} = \phi$}\\
    CBR_{\{e_1, \ldots, e_i\},c}\cap CBR_{e_{i+1}, c}, \text{ }~~~~~~\\ ~~~~~\mbox{otherwise}\\
    \end{array}
    \right.
\end{displaymath}
\noindent Finally, $CBR_{e,c} = CBR_{\{e_1, \ldots, e_m\}, c}$. It
is easy to verify that the CBRs of the entries generated by the
above approach satisfy Definition~\ref{def:cbre}.

For an entry $e$, similar to a user, we only store the CBRs of $e$
with respect to minimum degree constraints $2^0, 2^1, \cdots,$\hfill
$2^{\lfloor \log_2{c_e} \rfloor}$, where $c_e =max_{v\in V_e}{c_v}$
is the core number of $e$. Let $c_G$ denote the maximum core number
of the users in $G$ and $s$ denote the minimum fanout of an
SaR-tree. The total number of CBRs in an SaR-tree can be estimated
as,
\begin{align*} n_{CBR} &\leq \sum_{v\in V}{(\lfloor \log_2{c_v} \rfloor+1)} + \frac{2n(\lfloor \log_2{c_G} \rfloor+1)}{s} \\
&\leq n(\lfloor \log_2{\frac{\sum_{v\in V}{c_v}}{n}}\rfloor +
\frac{2(\lfloor \log_2{c_G} \rfloor+1)}{s} + 1).
\end{align*}
\noindent Since $c_G$ and $\frac{\sum_{v\in V}{c_v}}{n}$ are quite
small in a typical LBSN (e.g., they are 43 and 4.5 for the Gowalla
dataset used in our experiments), the storage cost of CBRs is
comparable to $G$ (e.g., around $2.3n$ in our experiments).


Based on the concept of CBRs, SaR-tree can be directly built on top
of R-tree. That is, we first construct a standard R-tree based on
the locations of the users and then embed the CBRs into each entry.
In this way, SaR-tree indexes both spatial locations and social
relations of the users. Note that the users in SaR-tree are
organized merely based on their locations --- they are spatially
close, but may not be well clustered in terms of their social relations.
This unfortunately weakens the pruning power of SaR-tree in processing GSGQs. To
overcome this weakness, we propose a variant in the next subsection.

\subsection{SaR*-tree} \label{isrtree}
Inspired by R*-tree, the R-tree variant that optimizes the grouping of spatial object to minimize the disk I/O cost, we propose SaR*-tree as an variant of SaR-tree. It has the same node structure but uses a different closeness metric to group users into nodes. Specifically, instead of using only the spatial area of MBR for closeness, SaR*-tree defines a new closeness metric $I(V)$ for a group of users $V$ that integrates both CBRs and MBRs to measure the combined social and spatial closenesses:
\begin{equation} I(V) = ||MBR_V|| \cdot \sum_{c}(||\cup_{v\in V}{CBR_{v,c}}-CBR_{V,c}||) \end{equation}
where $||\cdot||$ is the area of an MBR or CBR, 
and $\cup_{v\in V}{CBR_{v,c}}-CBR_{V,c}$ quantifies the similarity of CBRs of the users in $V$.
Obviously, a small $I(V)$ indicates that the users of $V$ have both
close locations and similar CBRs. This new closeness metric will be used in the R-tree construction.

Similar to SaR-tree, SaR*-tree is also constructed by
iteratively inserting users. During this construction, CBRs and MBRs
are generated at the same time and used for further user insertion.
Moreover, if a node $N$ of an SaR*-tree overflows, it will be split.
The details about these two main operations in SaR*-tree
construction, i.e., \emph{user insertion} and \emph{node split}, are
described below.
\begin{compactitem}
  \item \emph{User insertion}. When a user $v$ is inserted into an
SaR*-tree, for a node $N$ with entries $e_1, e_2, \cdots, e_m$, we
will select the entry $e_i$ with the minimal $I(V_{e_i} \cup \{v\})$
to insert $v$.
  \item \emph{Node split}. When a node $N$ of an SaR*-tree
overflows, we split $N$ into two sets of entries $N_1$ and $N_2$
with the minimal $I($ $\cup_{e_i\in N_1}V_{e_i})+I(\cup_{e_j\in
N_2}V_{e_j})$. Then, the parent node of $n$ use two entries to point
to $n_1$ and $n_2$, respectively. This splitting may propagate
upwards until the root.
\end{compactitem}





\section{GSGQ Processing} \label{app}
In this section, we present the detailed processing algorithms based
on SaR-trees for various GSGQs. As mentioned in
Section~\ref{problem}, we mainly focus on three types of GSGQs,
namely, $GSGQ_{range}$, $GSGQ_{rkNN}$, and $GSGQ_{kNN}$. We will
show that the CBRs of SaR-trees can be used in different ways for
processing these queries.


\subsection{GSGQ with Range Constraint}
When processing a $GSGQ_{range}$ $Q_{gs}=(v, range, c)$, each entry
of the SaR-tree or SaR*-tree that may cover result users will be
visited and possibly further explored. Compared to traditional
R-trees, which only provide spatial information via MBRs, an
SaR-tree or SaR*-tree provides much greater pruning power due to the
social information in CBRs. Consider an exemplary GSGQ $Q_{gs}=(v_1,
range, 2)$ in Fig.~\ref{fig:srtree}, where the shaded area is the
query range. When entry $b$ (which covers users $v_3$ and $v_4$) is
visited, $b$ needs further exploration if we only consider $MBR_{b}$
like in regular R-tree.
However, with $CBR_{b, 2}$, we can easily decide that any user group
inside the query range and containing any user in $V_{b}$ (i.e.,
$v_3$ or $v_4$), cannot be a $2$-core, because the query range is
covered by $CBR_{b, 2}$. Since $V_{b}$ does not contain any result
user, we can simply prune entry $b$ from further processing, as
formally proved in Theorem~\ref{the:range}. Considering SaR-trees
only maintain the CBRs with respect to exponential minimum degree
constraints, given a minimum degree $c$, we use $CBR_{v, 2^{\lfloor
\log_2{c} \rfloor}}$ to represent $CBR_{v, c}$ in $GSGQ_{range}$
processing. Similar ideas are also applied in $GSGQ_{rkNN}$ and
$GSGQ_{kNN}$ processing.

\begin{theorem}  \label{the:range}
For a $GSGQ_{range}$ $Q_{gs}=(v,
range, c)$ where $p_v\in range$, any user in $V_e$ of
entry $e$ does not belong to the result group if $range\subset CBR_{e,c}$
and $range$ does not contain any bounding edge of $CBR_{e,c}$.
\end{theorem}

\begin{proof}
We prove it by contradiction. If the theorem is not true, i.e., a
user $u\in V_e$ belongs to the result group $W$. Since the users of
$W\cup \{v\}$ are located inside $range$ and $range\subset
CBR_{e,c}$ does not contain any bounding edge of $CBR_{e,c}$, $W\cup
\{v\}$ is a $c$-core with $u$ inside $CBR_{e,c}$ (not including the
users on the bounding edges), which is contradictory to the CBR
definition for an entry.
\end{proof}

\begin{algorithm}
\caption{Processing $GSGQ_{range}$} \label{alg:pqrange}
\begin{algorithmic}[1]
\footnotesize
\REQUIRE LBSN $G=(V, E)$, $Q_{gs}=(v, range, c)$ \ENSURE Result of $Q_{gs}$\\
\emph{ProGSGQRange}($G$, $Q_{gs}$) \STATE Let $c'=2^{\lfloor
\log_2{c} \rfloor}$; \IF{$c_v < c$ or $range \subset CBR_{v,c'}$}
\STATE return $\phi$; \ENDIF \STATE Initialize $H$ with the root
entries of index tree; \WHILE{$H$ has non-leaf entries} \STATE Pop
the first non-leaf entry $e$ from $H$; \FOR{each child entry $e'$ of
$e$} \IF{$range \cap MBR_{e'} \neq \phi$ and $c_{e'}\geq c$ and
$range \not\subset CBR_{e',c'}$} \STATE Put $e'$ into $H$; \ENDIF
\ENDFOR \ENDWHILE \STATE Get the users $\widetilde{W}$ corresponding
to the entries of $H$; \STATE Compute the maximum $c$-core $W'$ of
$G[\widetilde{W}]$; \IF{$v\in W'$} \STATE return $W=W'-\{v\}$; \ELSE
\STATE return $\phi$; \ENDIF
\end{algorithmic}
\end{algorithm}

Algorithm~\ref{alg:pqrange} details the procedure of processing a
$GSGQ$ $_{range}$ based on an SaR-tree or SaR*-tree. At the beginning,
we access the CBR of user $v$. If $c_v < c$ or $range \subset
CBR$\hfill $_{v, 2^{\lfloor \log_2{c} \rfloor}}$, it means the core number of
$v$ is smaller than $c$ in the subgraph formed by the users inside
$range$. Thus, we cannot find any $c$-core containing $v$ inside
$range$ and there is no answer to $Q_{gs}$ (Lines 2-3). Otherwise,
we move on to find all candidate users $\widetilde{W}$ via the
proposed pruning schemes (Lines 6-13). Then, we compute the maximum
$c$-core $W'$ of $G[\widetilde{W}]$ by applying the
core-decomposition algorithm (Line 15). If $v\in W'$, $W = W' -
\{v\}$ is the answer; otherwise, there is no answer to $Q_{gs}$.

We again use the example in Fig.~\ref{fig:srtree} to illustrate the
pruning power of the proposed algorithm for processing
$GSGQ$$_{range}$. When applying the baseline algorithm based on
R-tree, $5$ users, i.e., $v_2$, $v_3$, $v_5$, $v_6$ and $v_8$, need
to be accessed. In contrast, in the proposed algorithm, by using
both MBRs and CBRs, there is no need to access index node $b$ (as
well as its covered user $v_3$) and user $v_8$ since $range\subset
CBR_{b, 2}$ and $range\subset CBR_{v_8, 2}$. As a result, only 3
users are accessed, achieving a great saving on computing and I/O
cost.

\subsection{GSGQ with Relaxed $k$NN Constraint}
\begin{figure}
\centering
\includegraphics[width=0.6\linewidth]{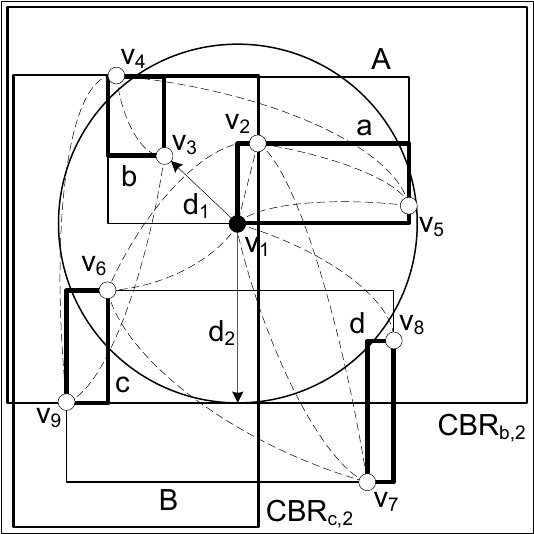}
\caption{An example of processing a $GSGQ_{rkNN}$ $Q_{gs}=(v_1, r3NN, 2)$.}
\label{fig:gsgknp} 
\end{figure}

To process a $GSGQ_{rkNN}$ $Q_{gs}=(v, rkNN, c)$ on an SaR-tree or
SaR*-tree, we maintain a priority queue $H$ of entries, whose priority score is the spatial distance from $v$ to both $MBR_e$ and
$CBR_{e,c}$. Let $L_{CBR_{e,c}}$ denote the set of bounding edges of
$CBR_{e,c}$ and $d(v,l)$ denote the distance from $v$ to edge $l$.
The distance from $v$ to $CBR_{e,c}$, where $v$ is located inside
$CBR_{e,c}$, is defined as the minimum distance from $v$ to reach
any bounding edge of $CBR_{e,c}$. Formally,
\begin{displaymath}d_{in}(v, CBR_{e,c})=\left\{
\begin{array}{ll}
    min_{l\in L_{CBR_{e,c}}}d(v, l),& \text{ } \mbox{$v\in CBR_{e,c}$}\\
    0, & \text{ } \mbox{otherwise}\\
    \end{array}
    \right.
\end{displaymath}
In our implementation, $d_{in}(v, CBR_{e,c})$ is computed based on
$CBR_{v, 2^{\lfloor \log_2{c} \rfloor}}$. $H$ uses $d_e =max \{d(v,
MBR_e), d_{in}(v,$\hfill$ CBR_{e,c})\}$ of an entry $e$ as the sorting key
in the queue. The rationale of adopting this priority queue is as
follows. By Definition~\ref{def:cbre} and the definition of
$d_{in}$, any user group inside the area $\odot(v, d_{in}(v,
CBR_{e,c}))$ and containing any user in $V_e$ cannot be a $c$-core.
In other words, if some users covered by entry $e$ belong to a
candidate group which satisfies the social acquaintance constraint,
the maximum distance of the candidate group to $v$ is expected to be
at least $d_e$. Therefore, we can derive another constraint on
$d_{max}(v, W)$ (recall that $d_{max}(v, W)$ is defined as $max_{u\in W}{d(v, u)}$) as summarized in Theorem~\ref{the:de} below. By
combing both constraints of $d_{max}(v, W)$ in $d_e$, we can get an
optimized processing order of the entries on an SaR-tree or
SaR*-tree. Fig.~\ref{fig:gsgknp} shows an example to demonstrate
this rationale. Suppose user $v_1$ issues a $GSGQ_{rkNN}$
$Q_{gs}=(v_1, r3NN, 2)$. When entry $b$ covering users $v_3$ and
$v_4$ is visited, we have $d_1 = d(v_1, MBR_{b})$ and $d_2 =
d_{in}(v_1, CBR_{b,2})$. Then, the key of $b$ is set to be $d_{b} =
max\{d_1, d_2\} = d_2$. We can see that if $v_3$ or $v_4$ belongs to
the result group, it should also contains $v_9$ to make the whole
group a $2$-core, which makes the maximum distance to $v_1$ larger
than $d_{c}$. Thus, we can access entry $c$ before $b$, although $c$
is spatially farther away from $v_1$ than $b$. As a result, a
candidate group $W=\{v_2, v_6\}$ can be obtained after accessing
entry $c$, since $d_{max}(v_1,W)<d_b$, there is no need to visit
entry $b$ any longer, thereby saving the access cost.


\begin{theorem}  \label{the:de}
Given a user $v$ and a minimum degree constraint $c$, if a user set
$W$ makes $G[W\cup \{v\}]$ a $c$-core, then $d_{max}(v, W)\geq d_e$
for any entry $e$ with $V_e\cap W \neq \phi$.
\end{theorem}


Algorithm~\ref{alg:pqknn} presents the details of processing a
$GSGQ$ $_{rkNN}$ based on an SaR-tree or SaR*-tree. A set
$\widetilde{W}$ is used to store the currently visited users and
initialized as $\{v\}$. The entries in $H$ are visited in ascending
order of $d_e$. If a visited entry $e$ is not a leaf entry, it will
be further explored and its child entries with $c_{e'}\geq c$ are
inserted into $H$ (Lines 7-10); otherwise, we get its corresponding
user $u$ (Line 12) and proceed with the following steps. If $c_u <
c$, it means $u$ cannot be a result user. Thus, we simply ignore it
and continue checking the next entry of $H$. On the other hand, if
$c_u \geq c$, $u$ is added into the candidate set $\widetilde{W}$
(Lines 13-14). Then, we compute the maximum $c$-core, denoted as
$W'$, in the subgraph formed by $\widetilde{W}$ (Line 15). If $|W'|
\geq k+1$ and $v\in W'$, $W'-\{v\}$ is the result (Line 16-17);
otherwise, the above procedure is continued until the result is
found or shown to be non-existent. Theorem~\ref{the:knn} proves the
correctness of Algorithm~\ref{alg:pqknn} and its superiority to the
baseline accessing model.

\begin{algorithm}
\caption{Processing $GSGQ_{rkNN}$} \label{alg:pqknn}
\begin{algorithmic}[1]
\footnotesize
\REQUIRE LBSN $G=(V, E)$, $Q_{gs}=(v, rkNN, c)$ \ENSURE Result of $Q_{gs}$\\
\emph{ProGSGQrKNN}($G$, $Q_{gs}$) \IF{$c_v < c$} \STATE return
$\phi$; \ENDIF \STATE $\widetilde{W} = \{v\}$; \STATE Initialize $H$
with the entries of the root node; \WHILE{$H \neq \phi$} \STATE
Pop the first entry $e$ from $H$; \IF{$e$ is not a leaf entry}
\FOR{each child entry $e'$ of $e$} \IF{$c_{e'} \geq c$} \STATE
Compute $d_{e'}$ and put $e'$ into $H$; \ENDIF \ENDFOR \ELSE \STATE
Get the corresponding user $u$ of $e$; \IF{$c_u \geq c$} \STATE
$\widetilde{W}=\widetilde{W}\cup \{u\}$; \IF{the first entry $e'$ in $H$ has $d_{e'} > d_e$} \STATE Compute the maximum
$c$-core $W'$ in $\widetilde{W}$; \IF{$|W'|\geq k+1$ and $v\in W'$}
\STATE return $W' - \{v\}$; \ENDIF \ENDIF \ENDIF \ENDIF \ENDWHILE \STATE
return $\phi$;
\end{algorithmic}
\end{algorithm}

\begin{theorem}  \label{the:knn}
For a $GSGQ_{rkNN}$ $Q_{gs}=(v, rkNN, c)$, Algorithm~\ref{alg:pqknn}
generates the result of $Q_{gs}$. Moreover, it checks equal or less
users than that of the baseline accessing model based on $d(v,
MBR_e)$.
\end{theorem}

\begin{proof}
Let $W$ be the user set returned by Algorithm~\ref{alg:pqknn}. %
and user $u' = \arg_{u\in W}{\max{d(v,u)}}$. Suppose another user
set $W'$, $W'\neq W$, is the result. Then, it should be either 1) $d_{max}(v, W') < d_{max}(v, $$W)$ or
2) $d_{max}(v, W') = d_{max}(v, W)$ and $W \subset W'$.

For case 1), consider a user $u\in W'$. For any entry $e$ which covers $u$, based on Theorem~\ref{the:de}, we
have $d_e \leq d_{max}(v, $ $W') < d_{max}(v, W) = d(v, u’) \leq d_{u'}$.
According to Algorithm~\ref{alg:pqknn}, a super set of $W'$, denoted
as $W''$, should be checked before getting $W$ and $G[W''\cup
\{v\}]$ does not contain a $c$-core of size no less than $k+1$ covering $v$. Then, $W'$ cannot
be the result, which is contradictory to the assumption.

For case 2), consider a user $u\in W'$ and $u\notin W$. According to Algorithm~\ref{alg:pqknn},
there is a any entry $e$ which covers $u$ and $d_e > d_{u’}$. Based on Theorem~\ref{the:de}, we
have $d_{max}(v, W') \geq d_e > d_{u'} \geq d(v, u’) = d_{max}(v, W)$, which is contradictory to the assumption $d_{max}(v, W') = d_{max}(v, W)$.

To conclude, $W$ is the result of $Q_{gs}$.

Let $S$ and $S'$ be the entries explored by
Algorithm~\ref{alg:pqknn} and by the baseline accessing model based
on $d(v, MBR_e)$, respectively, for finding the result set $W$.
Based on Theorem~\ref{the:de}, for any entry $e\in S$, we have
$d_e\leq d_{max}(v, W)$ (if not, $e$ will not be further explored
since all the users of $W$ have been accessed and the result set $W$
has been found). Considering that $S'=\{e|d(v, MBR_e)\leq d_{max}(v,
W)\}$, then $S\subseteq \{e|e\in S' \wedge d_{in}(v, CBR_{e,c})\leq
d_{max}(v, W)\}\subseteq S'$. It means $S$ contains equal or less
users than that of $S'$.
\end{proof}

Recall the example in Fig.~\ref{fig:gsgknp}. When applying
Algorithm~\ref{alg:pqknn} to process $Q_{gs}=(v_1, r3NN, 2)$, the
access order of the users is $v_2$, $v_6$, $v_5$, $v_3$, $v_4$,
$v_9$ and $v_7$. The result can be obtained by accessing the first
$3$ users. In contrast, the baseline algorithm based on R-tree
accesses the users in the order of $v_2$, $v_3$, $v_6$, $v_5$,
$v_4$, $v_8$, $v_9$, and $v_7$. Then, $4$ users are accessed and
processed. Obviously, by reorganizing the access order of entries,
Algorithm~\ref{alg:pqknn} processes $GSGQ_{rknn}$ more efficiently.

\subsection{GSGQ with Strict $k$NN Constraint} \label{app:knn}
For a $GSGQ_{kNN}$ $Q_{gs}=(v, kNN, c)$, we adopt the same
processing framework as in Algorithm~\ref{alg:pqknn}. However, when
a valid $W'$ is found for $GSGQ_{rkNN}$ at Line 16, more steps will
be needed to obtain the result of $GSGQ_{kNN}$. Let $W'$ be the
maximum $c$-core formed by the set of currently visited users
$\widetilde{W}$. Only if $|W'|\geq k+1$ and $v\in W'$, it is
possible to find a $c$-core of size $k + 1$ in $\widetilde{W}$ that
contains $v$. Moreover, such a $c$-core must be a subset of $W'$.
Thus, we invoke a function \emph{FindExactkNN} to check all user
sets of size $k+1$ that contain $v$ in $W'$. If such a user set
$W''$ is found, $W'' - \{v\}$ is the result of $Q_{gs}$; otherwise,
the above procedure is repeated when Algorithm~\ref{alg:pqknn}
continues to find the next candidate $W'$.

\textbf{In-Memory Optimizations.} The above processing framework
provides optimized node access on SaR-trees for $GSGQ$ $_{kNN}$.
However, due to the NP-hardness of $GSGQ_{kNN}$, the in-memory
processing function \emph{FindExactkNN} also has a great impact on
the performance of the algorithm. A naive idea of checking all
possible combinations of the user sets costs up to exponential time
complexity of $k$. In this subsection, we single out this problem to
optimize the \emph{FindExactkNN} function by designing two pruning
strategies.

Algorithm~\ref{alg:findexactknn} details the optimized
\emph{FindExactkNN}, which employs a \emph{branch-and-bound} method
and expands the source user set $S$ from the candidate user set $U$.
At the beginning, $S$ and $U$ are initialized as $\{v, u\}$ and $W'
-\{v, u\}$ ($u$ denotes the newly accessed user in
Algorithm~\ref{alg:pqknn}), respectively. Note that if a result
$W''$ exists in $W'$, $W''$ must contain $u$, because it has been
proved that $W'-\{u\}$ does not contain a result. During the
processing, two major pruning strategies, namely,
\emph{core-decomposition based pruning} (Lines 6-11, 16-20) and
\emph{$k$-plex based pruning} (Lines 5, 12), are applied.

\textit{1) Core-Decomposition based Pruning:} Based on the
definition of $c$-core, we can observe that if the current source
user set $S'$ can be expanded to a $c$-core of size $k+1$, it must
be contained by the maximum $c$-core of $U'\cup S'$, where $U'$
denotes the set of remaining candidate users. Therefore, we conduct
a core-decomposition on $U'\cup S'$ before further exploration. If a
user of $S'$ has a core number smaller than $c$ in $U'\cup S'$, $S'$
cannot be expanded to a result from the candidate user set $U'$ and
thus we can safely stop further exploration. In addition, if the
maximum $c$-core in $U'\cup S'$ contains $S'$ and has size $k+1$, it
is the result of $GSGQ_{kNN}$ and the whole processing terminates.
Otherwise, further exploration on the maximum $c$-core of $U'\cup
S'$ is required. Finally, if $S'$ cannot be expanded to a $c$-core
of size $k+1$, we roll back to explore $S$ and the remaining $U$.
Similarly, we compute the maximum $c$-core $W'$ of $S\cup U$. If
$|W'| \geq k+1$ and $S\subseteq W'$, $S$ could be expanded to the
result from $U=W'-S$ and further exploration is applied; otherwise,
no result can be found.

\textit{2) $k$-plex based Pruning:} One major challenge of the
$c$-core problem is that it does not preserve locality, that is, if
$W$ is a $c$-core, adding or dropping some users
from $W$ no longer retains it as a $c$-core. 
As a workaround, we transfer the problem to a dual $\bar{c}$-plex problem
\cite{balas}  (which preserves the locality property) by
adding some constraint. Simply speaking, a $\bar{c}$-plex
$W\subseteq V$ is a set such that $\delta(G[W]) \geq |W| - \bar{c}$.

Since a $c$-core of size $k + 1$ is also a $(k+1-c)$-plex, we seek
to find a $(k+1-c)$-plex of size $k + 1$ to achieve further pruning.
$\bar{c}$-plex preserves the locality property because if $W$ is a
$\bar{c}$-plex, dropping some users can still make it a
$\bar{c}$-plex. In other words, if the maximum $(k+1-c)$-plex in
$U'\cup S'$ has a size no less than $k+1$, it is certain that a
$(k+1-c)$-plex of size $k+1$ can be found; otherwise, such a
$(k+1-c)$-plex cannot be found. Moveover, $(k+1-c)$-plex is more
constrained than $c$-core because the size of the maximum
$(k+1-c)$-plex is always no larger than that of the maximum $c$-core
of size no smaller than $k+1$.

\begin{algorithm}
\caption{Finding $c$-core of size $k+1$} \label{alg:findexactknn}
\begin{algorithmic}[1]
\footnotesize
\REQUIRE User set $U$ and $S$, $c$, $k$ \ENSURE $c$-core $W$ \\
\emph{FindExactkNN}($U$, $S$, $c$, $k$) \IF{$|S| = k+1$} \STATE
return $S$; \ENDIF \WHILE{$U\neq \phi$} \STATE $S' = S \cup \{u\}$,
$U=U - \{u\}$ for some $u\in U$; \STATE $U' = \{u\in U: S' \cup
\{u\}$ is a $(k+1-c)$-plex $\}$; \STATE Compute the maximum $c$-core
$W'$ of $U'\cup S'$; \IF{$|W'| \geq k+1$ and $S'\subseteq W'$}
\IF{$|W'|=k+1$} \STATE return $W'$; \ELSE \STATE $U'=W'-S'$;
\IF{$B_p(G[U'\cup S']) \geq k+1$} \STATE $W''$ =
\emph{FindExactkNN}($U'$, $S'$, $c$, $k$); \IF{$W'' \neq \phi$}
\STATE return $W''$; \ENDIF \ENDIF \ENDIF \ENDIF \STATE Compute the
maximum $c$-core $W'$ of $S\cup U$; \IF{$|W'| \geq k+1$ and
$S\subseteq W'$} \STATE $U = W' - S$; \ELSE \STATE break; \ENDIF
\ENDWHILE \STATE return $\phi$;
\end{algorithmic}
\end{algorithm}

The properties of $(k+1-c)$-plex can be used to devise powerful
pruning strategies in processing $GSGQ_{kNN}$. First, we prune those
users in $U$ who cannot expand the source user set $S'$ to a
$(k+1-c)$-plex. This pruning is implemented in Line 5 of
Algorithm~\ref{alg:findexactknn}. Second, we estimate the size of a
maximum $(k+1-c)$-plex to provide further pruning. Some theoretic
bounds on it have been proposed in the literature. In this paper, we
adopt the result of \cite{mcclosky} and compute an upper bound $B$
on the size of a maximum $(k+1-c)$-plex in a graph $G$ as,
\begin{equation} B_p(G) = min_{i= 1, \ldots, p}\{\frac{1}{i} B(C^i_1, \ldots, C^i_{m_i})\}, \end{equation}
and
\begin{align*} B(C^i_1, \ldots, C^i_{m_i}) = &\sum_{j = 1}^{m_i}min\{2\bar{c} - 2 + \bar{c}\ mod\ 2,
\bar{c}+a_{i,j}, \\ &\Delta(G[C^i_j])+\bar{c}, |C^i_j|\},
\end{align*} where $\bar{c} = k+1-c$, $C^i_{1}, \ldots, C^i_{m_i}$
are co-$\bar{c}$-plexes \cite{mcclosky} in which every vertex of $V$
appears exactly $i$ times, $a_{i,j} = max\{n: |\{v|v \in V \wedge
deg_G(v) \geq n\}| \geq \bar{c} + l\}$ for each $C^i_j$, and $p$ is
a parameter to limit the iterations of computing.

\begin{figure}
\centering
\includegraphics[width=\linewidth]{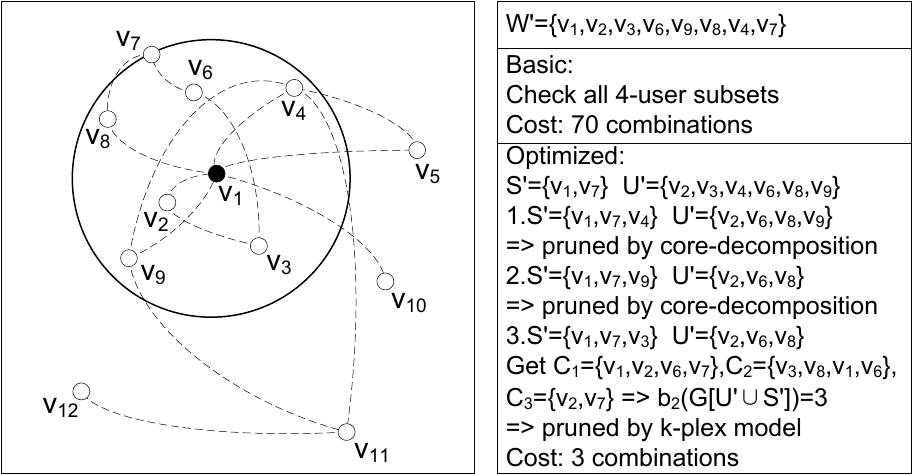}
\caption{Exemplary procedures of the original and optimized function
\emph{FindExactkNN} when $W' =\{v_1, v_2, v_3, v_6, v_9, v_8, v_4,
v_7\}$ for $GSGQ_{kNN}$ $Q_{gs}=(v_1, 3NN, 2)$. The entries are
omitted here because they are not related to function
\emph{FindExactkNN}. } \label{fig:kpopt} 
\end{figure}

Fig.~\ref{fig:kpopt} shows the steps of both the basic and optimized
version of function \emph{FindExactkNN} where user set $W' =\{v_1,
v_2, v_3, v_6, v_9, v_8,$ $v_4, v_7\}$ and $Q_{gs}=(v_1, 3NN, 2)$.
In the optimized procedure, each step shows the investigated source
user set $S'$ and the candidate set $U'$ after filtering. For
example, in the first step, we try to check $S'=\{v_1, v_7, v_4\}$
and $U'=\{v_2, v_3, v_6, v_8, v_9\}$. After filtering $U'$ via Line
5 of Algorithm~\ref{alg:findexactknn}, we can get $U'=\{v_2, v_6,
v_8, v_9\}$. Since the maximum $2$-core of $U'\cup S'$ only has size
$3$, no $2$-core of size $4$ can be found in $U'\cup S'$. Thus, all
the combinations of these users can be ignored. A similar case can
be found in the second step when $S'=\{v_1, v_7, v_9\}$. In the
third step, we can get the upper bound of the size of the maximum
$2$-plex in $U'\cup S'$ as $3$ by computing $B_2(G[U'\cup S'])$.
Thus, $U'\cup S'$ does not contain a $2$-core of size $4$. We can
stop searching here because no user is filtered from $U'$ in the
last step, which means all the combinations are covered. We can see
that the optimized function \emph{FindExactkNN} effectively prunes
unnecessary explorations and saves significant computation cost.

\section{Update of SaR-trees} \label{update}
The SaR-trees, once built, can be used as underlying structures for
efficient GSGQ processing with generic spatial constraints. It is
particularly favorable for applications where both social relations
and user locations (e.g., home addresses) are stable. However, for
other applications where users may regularly change their locations
and social relations, efficient update of the SaR-trees is required.
This is challenging because an update of a user affects not only her
own CBR but also those of others. In this section, we propose a lazy
update approach tailored for SaR-trees that strikes a balance between update
efficiency and effectiveness of GSGQ processing. 

\subsection{Lazy Update in SaR-trees}
An update from user $v\in G$ means either her location changes from
$p_v$ to $p'_v$ or her social relation $N_G(v)$ changes. However,
not all changes lead to the update of CBRs. The following
two rules show the location and social conditions on which CBRs might need updates.

\begin{updaterule} \label{pro:loc}
{\bf Location update.} A $CBR_{u,c}$ might become invalid only if
there exists some user $v$ such that $c \leq c_v$, $p_v\notin
CBR_{u,c}$, and $p'_v\in CBR_{u,c}$.
\end{updaterule}

\begin{updaterule} \label{pro:soc}
{\bf Social updates.} A $CBR_{u,c}$ might become invalid only if there
exist two users $v, v'$ such that edge $vv'$ is newly added,
$min\{c_v, c_{v'}\} \geq c$ and $\{p_v, p_{v'}\}\in CBR_{u,c}$.
\end{updaterule}

\begin{figure}
\centering
\includegraphics[width=\linewidth]{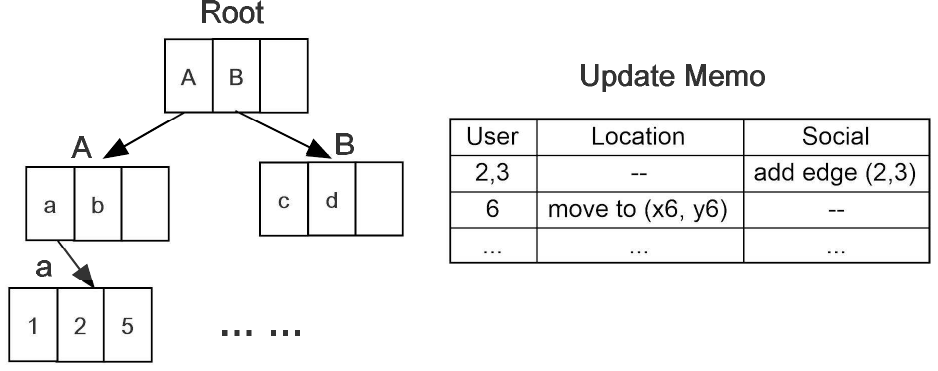}
\caption{Lazy Update and Update Memo}
\label{fig:lazyupdate} 
\end{figure}

To relieve an update procedure from intensive CBR re-computation, we
propose a lazy update model for SaR-trees. Particularly, a memo $M$
is introduced to store those accumulated updates which have not been
applied on the CBRs of SaR-trees. Fig.~\ref{fig:lazyupdate} illustrates
the data structure for the SaR-tree in Fig.~\ref{fig:srtree}. A user
update is thus handled in three steps. In the first step, the user
record is updated, and core-decomposition is performed on $G$ to
update the core numbers of users if it is a social update. If the
core number of a user $u$ changes, the core numbers of the entries
along the path from $u$ to the root are updated. In the second step,
the user update is added into $M$. In this figure, user $v_2$ adds
an edge with $v_3$, and the new edge has been inserted to $M$.
Similar operation is performed for location updates when a user
moves into other users' CBRs. In the third step, when the size of
$M$ reaches a threshold, named the \emph{Batch Update Size}, a batch
update is applied on the CBRs of SaR-trees. This calls for
re-computation of all affected CBRs in $M$.

To facilitate CBR updates, an R-tree is built on the CBRs of users.
By a point containment query on this R-tree, we can find the CBRs
that cover the latest location of an updated user. The retrieved
CBRs are then filtered based on Update Rule~\ref{pro:loc} and Update
Rule~\ref{pro:soc}. For the remaining CBRs, we first determine their
validity by computing the core numbers of the corresponding users in
the subgraphs formed by the users inside the CBRs. Then, each
invalid CBR is recomputed by applying Algorithm~\ref{alg:compcbr}
and its update is propagated to the root along the SaR-tree path.

\subsection{GSGQ Processing with Update-Memo on SaR-trees}
With an update-memo $M$, GSGQ processing algorithms on SaR-trees
need to be revised for correctness as some CBRs may be invalid. In
the following, we outline the major changes of the processing
algorithms for different GSGQs.

\textbf{$GSGQ_{range}$ processing.} To revise
Algorithm~\ref{alg:pqrange}, the CBRs will no longer be used to
prune entries when traversing the SaR-tree. As a result, the
priority queue $H$ is composed of a number of leaf entries, each
corresponding to a user with core number equal to or larger than $c$
inside $range$. As such, for each user $u$ in $H$ s.t. $range\subset
CBR_{u,c}$, we check the other users in $H$ located inside
\emph{range}: if some other user has updates in $M$ which might
invalidate $CBR_{u,c}$ according to Update Rule~\ref{pro:loc}
or~\ref{pro:soc}, we keep $u$ in $H$; otherwise, $u$ is pruned from
$H$. In the end, if the query issuer $v$ is pruned from $H$, there
will be no result; otherwise, we obtain the result from $H$ as
Algorithm~\ref{alg:pqrange} does.

\textbf{$GSGQ_{rkNN}$ (or $GSGQ_{kNN}$) processing.} To revise
Algorithm~\ref{alg:pqknn}, we still use the second priority queue
$H'$ to store the entries of $H$ in ascending order of their minimal
distances to $v$. When putting an entry $e$ into $H$, if $d_{in}(v,
CBR_{e,c})$ $> d(v, MBR_e)$, we need to verify the validity of
$CBR_{e,c}$. For a non-leaf entry $e$, we simply set $d_e=d(v,
MBR_e)$ to avoid the validating cost. For a leaf entry $e$, let $u$
be the corresponding user. We retrieve all users with shorter
distances to the query issuer $v$ than $d_{in}(v, CBR_{e,c})$ by
exploring $H'$, denoted as $U$. Then, we filter out the users in $U$
who has no update in $M$ or cannot invalidate $CBR_{e,c}$ according
to Update Rule~\ref{pro:loc} or~\ref{pro:soc}. If $U$ is not empty,
$d_{in}(v, CBR_{e,c})$ is updated as $min_{u'\in U}d(v, u')$. It is
easy to verify that if $p_u\in \odot(v, min_{u'\in U}d(v, u'))$, any
user group with $u$ inside $\odot(v, min_{u'\in U}d(v, u'))$ cannot
be a $c$-core. This guarantees the correctness of the algorithm.

\section{Performance Evaluation} \label{experiment}
In this section, we evaluate the proposed methods on three real
datasets, namely, \emph{Gowalla}, \emph{Dianping}, and \emph{Twitter-2010}, and
investigate the impact of various parameters. The code is written in
C++ and compiled by GNU gcc x64 4.5.2. All the experiments are
performed on a Dell R430 server with dual Intel Xeon E5-2620 CPU and 64GB RAM, running GNU/Ubuntu Linux 64-bit 14.04 LTS.

\subsection{Experimental Setting}
The Gowalla dataset was collected from the location-based social network Gowalla (available on \url{http://snap.stan-ford.edu/data/loc-gowalla.html}), the Dianping
dataset was crawled by us from a Chinese restaurant review site (available on \url{https://goo.gl/uUV4Wg}), and the Twitter-2010 dataset is from the social network Twitter (available on
\url{http://law.di.unimi.it/webdata/twitter-2010/}). For the Gowalla dataset and the Dianping dataset, we remove the users with no
check-ins and select the first check-in position of each user
as his/her location. As a result, the preprocessed Gowalla
dataset has 107,092 nodes (users) and 456,830 edges (friend relations), while the preprocessed Dianping dataset has 2,673,970 nodes and 922,977 edges.
In comparison, the Twitter-2010 dataset is much bigger, with 41,652,098 nodes and 684,500,219 edges. The locations of the users in Twitter-2010 are randomly
generated with a uniform distribution. For both datasets, we normalize the location data into a unit space [0,1] x [0, 1].

We implement four indexes for performance evaluation, namely,
\emph{R-tree}, \emph{C-imbedded R-tree}, \emph{SaR-tree}, and
\emph{SaR*-tree}. The C-imbedded R-tree is built on top of an R-tree and additionally stores the core numbers of the index entries.
The average CPU time of constructing a user CBR in the latter
two trees is less than $100$ ms for Gowalla and Dianping, and $50$ ms for
Twitter-2010. The sizes of SaR-trees are 15.5MB for Gowalla, 257MB for
Dianping and
2.1GB for Twitter-2010. The index construction time is less
than 1 minute for Gowalla and Dianping, and 1.3 hours for Twitter-2010. The
corresponding GSGQ processing methods on these
indexes are denoted as \emph{BR} (baseline R-tree), \emph{CR},
\emph{SaR} and \emph{SaR*}, respectively. \emph{CR} enhances
\emph{BR} by pruning those nodes whose core numbers cannot
satisfy the minimum degree constraint $c$ in query processing.

\begin{table}
\centering \caption{System parameter settings}
\label{tab:parameters}
\begin{tabular}{|l|l||l|l|} \hline
Parameter & Value & Parameter & Value \\ \hline $c$ &
$1-5$ & $r$ & $0.002-0.05$ \\ \hline  $k$ & $10-250$ & Page size & 4KB \\ \hline Page acc. time & 2ms & &\\
\hline \multicolumn{4}{|c|}{Gowalla}
\\ \hline User \# & $107,092$ & Edge \# & $456,830$ \\ \hline Max degree & $9,967$ & Avg. degree & $9.177$ \\
\hline Max core num. & $43$ & Avg. core num. & $4.839$ \\
\hline Dataset size & 27.2MB & & \\
\hline \multicolumn{4}{|c|}{Dianping} \\ \hline User \# & $2,673,970$ &
Edge \# & $922,977$ \\ \hline Max degree & $11423$ & Avg. degree & $5.184$ \\
\hline Max core num. & $24$ & Avg. core num. & $2.741$ \\
\hline Data size & 162M & & \\
\hline \multicolumn{4}{|c|}{Twitter-2010} \\ \hline User \# &
$41,652,098$ & Edge \# & $684,500,219$ \\ \hline Max degree & $1,405,986$ & Avg. degree & $30.453$ \\
\hline Max core num. & $2,059$ & Avg. core num. & $14.692$ \\
\hline Dataset size & 29.7GB & & \\
\hline
\end{tabular}
\end{table}

To have a fair comparison, we implement CR, SaR, and SaR* by
coupling extra pages with each index node to store the information
of core numbers (for CR) or CBRs (for SaR and SaR*). These extra
pages are called \emph{coupled nodes}. To compare the
performance of different methods, we mainly use two metrics, namely, the
page access cost and the query running time. The former includes the
page accesses of index nodes, coupled nodes, and user data. On the
other hand, the query running time measures the actual clock time to
process a GSGQ, including the CPU time and the I/O time. In the
experiments, no cache is used for GSGQ processing and the page
access time is set as $2$ ms per page access. Each test ran a set of 1,000 randomly
generated GSGQs and we report the average performance.

Three types of queries, namely, $GSGQ_{range}$, $GSGQ_{rkNN}$, and
$GSGQ_{kNN}$, are tested. For $GSGQ_{range}$, the range $r$ is
defined as a square centered at the location of the query issuer.
In the sequel, we use the edge length to represent
$r$, which is set at $0.002$ for Gowalla and Dianping, and $0.05$ for Twitter-2010
by default. For $GSGQ_{rkNN}$ and $GSGQ_{kNN}$, $k$ is selected from
$10$ to $250$, which represents large-scale time-consuming queries for
real-life social applications, e.g., the marketing example shown in
Section~\ref{sec:intro}. Finally, the minimum degree constraint $c$
is selected from $1$ to $5$. Table~\ref{tab:parameters} summarizes
the major parameters and their values used in the experiments, where
the average degree only counts connected nodes.

\subsection{Overall Performance}\label{sec:overall}
Table~\ref{tab:effect} shows the average minimum degree of the
result groups for three different query semantics on Gowalla, where
$kNN$ denotes a classic $k$-nearest-neighbor query and $SSGQ$
denotes the socio-spatial group query proposed in \cite{yang12}. As
expected, GSGQ always retrieves the groups that satisfy the minimum
degree constraints, while the other two queries have a minimum degree of close to zero. This justifies the improved social
constraint introduced by GSGQ.

\begin{table}
\centering \caption{Minimum degree of the result group given $k=50$
on Gowalla.} \label{tab:effect}
\begin{tabular}{|l|c|c|c|c|c|} \hline
\multirow{2}{*}{Query} & \multicolumn{5}{c|}{$\rho$} \\
\hhline{~-----} & 1 & 2 & 3 & 4 & 5 \\
\hline $kNN$ & 0 & 0 & 0 & 0 & 0 \\
\hline $SSGQ (p=\rho)$ & 0.05 & 0.08 & 0.11 & 0.16 & 0.21 \\
\hline $GSGQ_{rkNN} (c= \rho)$ & 1 & 2 & 3 & 4 & 5 \\
\hline
\end{tabular}
\end{table}
\begin{figure}
  \centering
  \subfigure[$GSGQ_{range}$ (r=0.002, c=4)]{
    \label{fig:subfig:rangelatency} 
    \includegraphics[width=0.48\linewidth]{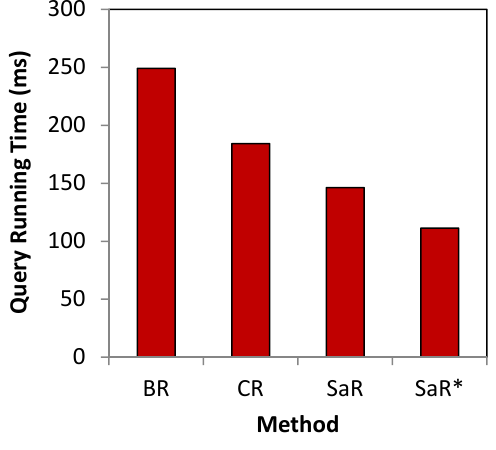}}
  \subfigure[$GSGQ_{range}$ (r=0.002, c=4)]{
    \label{fig:subfig:rangeio} 
    \includegraphics[width=0.48\linewidth]{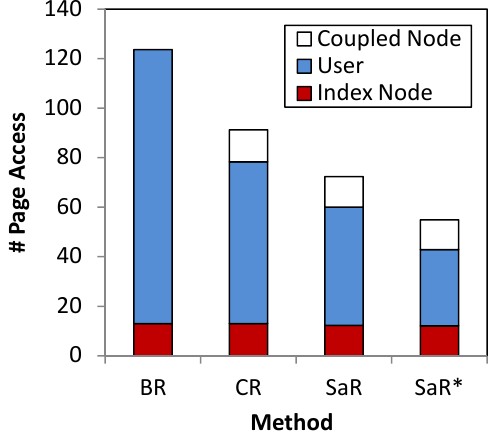}}
  \subfigure[$GSGQ_{rkNN}$ (k=100, c=4)]{
    \label{fig:subfig:knnlatency} 
    \includegraphics[width=0.48\linewidth]{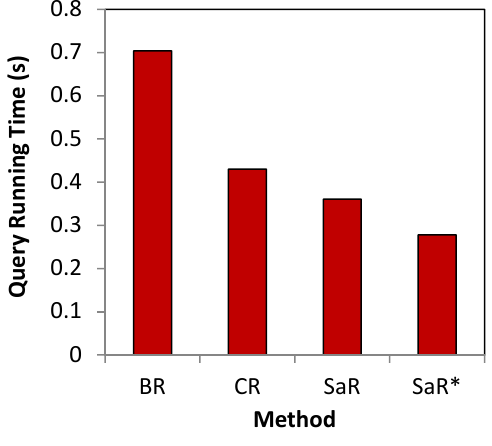}}
  \subfigure[$GSGQ_{rkNN}$ (k=100, c=4)]{
    \label{fig:subfig:knnio} 
    \includegraphics[width=0.48\linewidth]{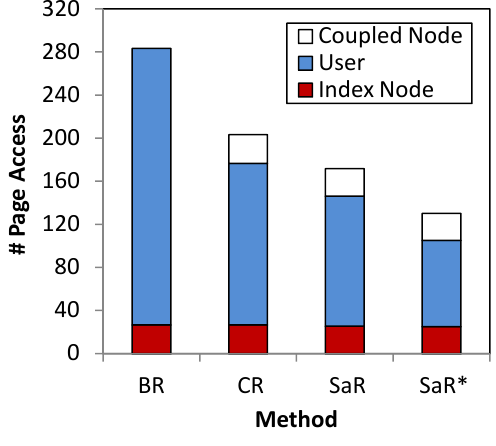}}
  \subfigure[$GSGQ_{kNN}$ (k=100, c=3)]{
    \label{fig:subfig:knnelatency} 
    \includegraphics[width=0.48\linewidth]{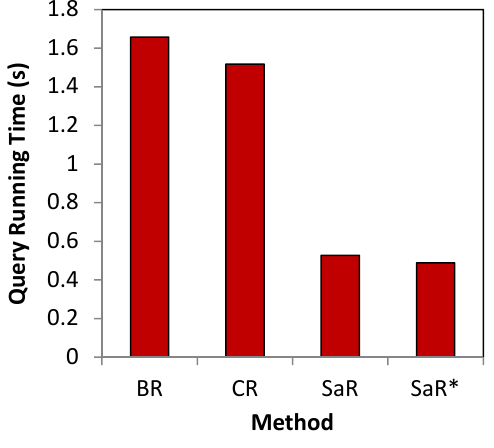}}
  \subfigure[$GSGQ_{kNN}$ (k=100, c=3)]{
    \label{fig:subfig:knneio} 
    \includegraphics[width=0.48\linewidth]{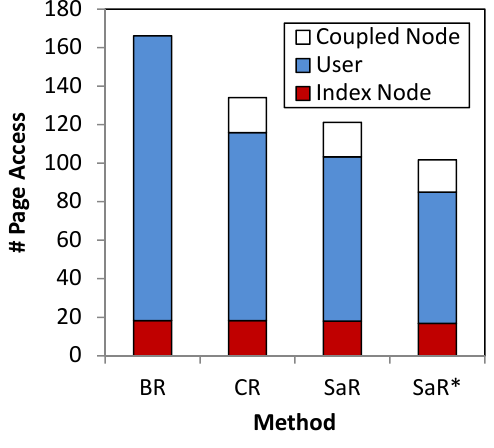}}
  \caption{Overall performance comparison on Gowalla.}
  \label{fig:overallperform} 
\end{figure}

\begin{figure}
  \centering
  \subfigure[$GSGQ_{range}$ (r=0.002, c=4)]{
    \label{fig:subfig:dprangelatency} 
    \includegraphics[width=0.48\linewidth]{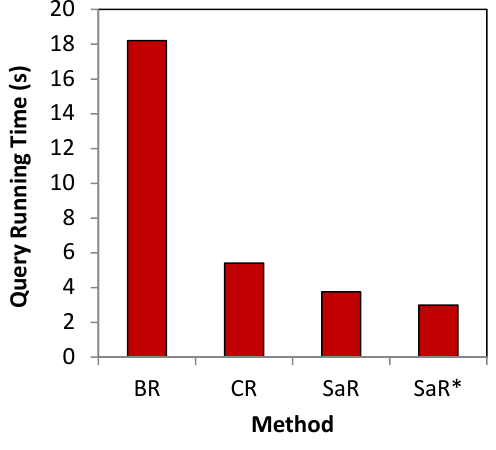}}
  \subfigure[$GSGQ_{range}$ (r=0.002, c=4)]{
    \label{fig:subfig:dprangeio} 
    \includegraphics[width=0.48\linewidth]{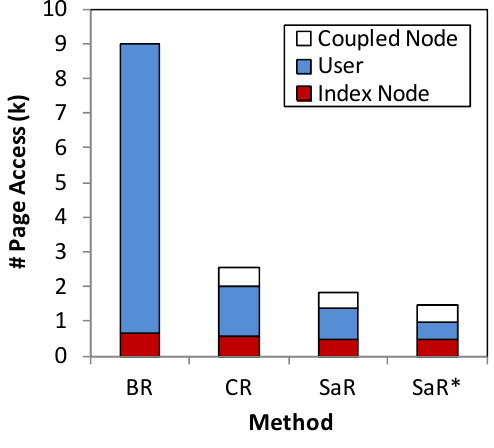}}
  \subfigure[$GSGQ_{rkNN}$ (k=100, c=4)]{
    \label{fig:subfig:dpknnlatency} 
    \includegraphics[width=0.48\linewidth]{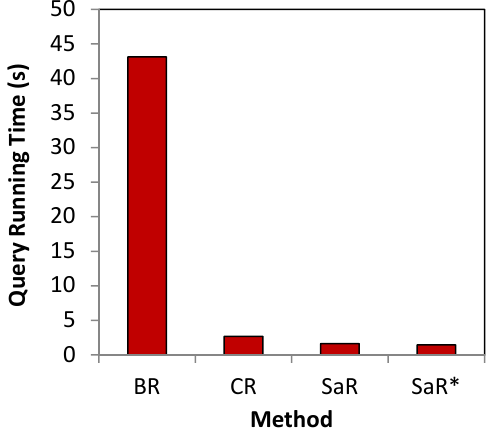}}
  \subfigure[$GSGQ_{rkNN}$ (k=100, c=4)]{
    \label{fig:subfig:dpknnio} 
    \includegraphics[width=0.48\linewidth]{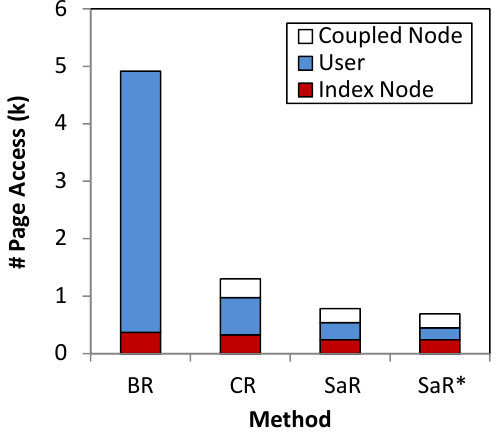}}
  \subfigure[$GSGQ_{kNN}$ (k=30, c=8)]{
    \label{fig:subfig:dpknnelatency} 
    \includegraphics[width=0.48\linewidth]{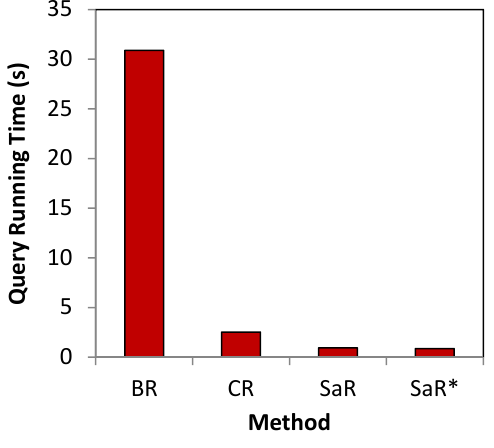}}
  \subfigure[$GSGQ_{kNN}$ (k=30, c=8)]{
    \label{fig:subfig:dpknneio} 
    \includegraphics[width=0.48\linewidth]{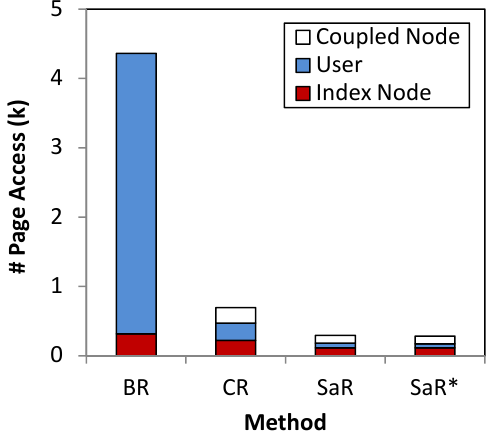}}
  \caption{Overall performance comparison on Dianping.}
  \label{fig:dpoverallperform} 
\end{figure}

\begin{figure}
  \centering
  \subfigure[$GSGQ_{range}$ (r=0.05, c=2)]{
    \label{fig:subfig:twrangelatency} 
    \includegraphics[width=0.48\linewidth]{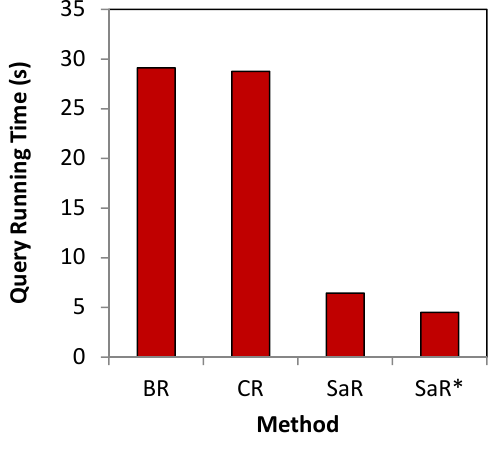}}
  \subfigure[$GSGQ_{range}$ (r=0.05, c=2)]{
    \label{fig:subfig:twrangeio} 
    \includegraphics[width=0.48\linewidth]{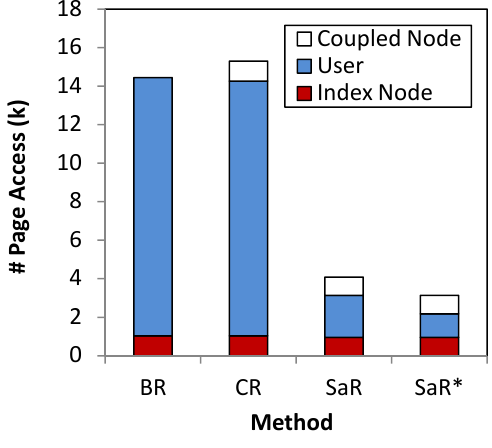}}
  \subfigure[$GSGQ_{rkNN}$ (k=20, c=2)]{
    \label{fig:subfig:twknnlatency} 
    \includegraphics[width=0.48\linewidth]{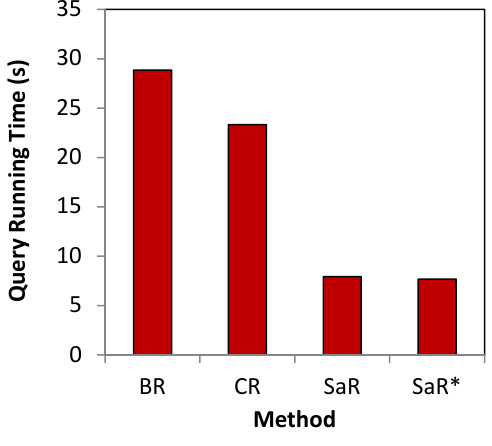}}
  \subfigure[$GSGQ_{rkNN}$ (k=20, c=2)]{
    \label{fig:subfig:twknnio} 
    \includegraphics[width=0.48\linewidth]{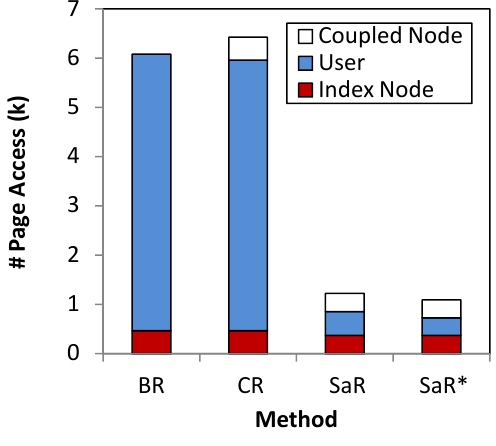}}
  \subfigure[$GSGQ_{kNN}$ (k=20, c=2)]{
    \label{fig:subfig:twknnelatency} 
    \includegraphics[width=0.48\linewidth]{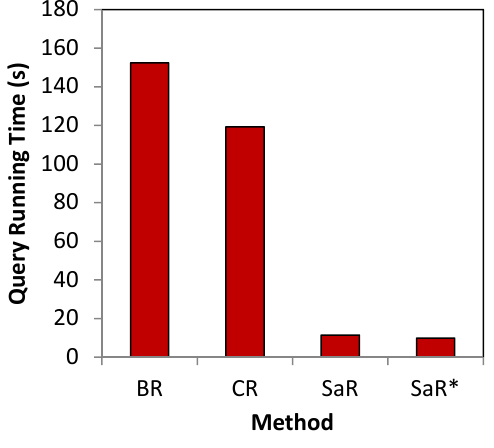}}
  \subfigure[$GSGQ_{kNN}$ (k=20, c=2)]{
    \label{fig:subfig:twknneio} 
    \includegraphics[width=0.48\linewidth]{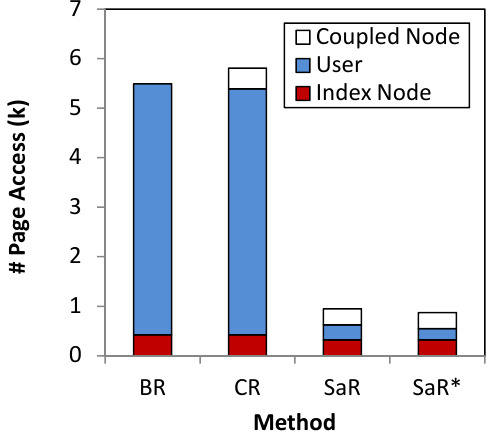}}
  \caption{Overall performance comparison on Twitter-2010.}
  \label{fig:twoverallperform} 
\end{figure}

Fig.~\ref{fig:overallperform}, Fig.~\ref{fig:dpoverallperform} and Fig.~\ref{fig:twoverallperform}
show the overall performance of the GSGQ methods under three
different queries on Gowalla, Dianping and Twitter-2010, respectively. Generally,
SaR and SaR* achieve significant improvement over BR and CR
in all tested cases. Take Twitter-2010 as an example. For $GSGQ_{range}$,
SaR and SaR* outperform BR and CR by $77.9\% - 77.6\%$ and $84.5\% - 84.3\%$ in
terms of the query running time (see
Fig.~\ref{fig:subfig:twrangelatency}). This is mainly due to the
savings in accessing the user data as shown in
Fig.~\ref{fig:subfig:twrangeio}. It is interesting to note that CR incurs an even higher page access cost than BR because
of the week pruning power of the core numbers for large social networks and
additional accesses on the coupled nodes. More specifically, SaR and SaR* check
much fewer users (around 2,946 users) than CR (around 103,060 users)
and BR (around 85,686 users) to derive the results. SaR* further
reduces the page accesses to 3,135 compared to SaR (4,089), CR (15,293),
and BR (14,436). All the results exhibit the high pruning power of CBRs
for $GSGQ_{range}$ processing.

For $GSGQ_{rkNN}$, SaR and SaR* achieve similar improvement
over BR and CR in terms of the query running time and the page access cost (see Fig.~\ref{fig:subfig:twknnlatency} and
Fig.~\ref{fig:subfig:twknnio}). They access much less users in query processing. Specifically, SaR and SaR* only
check $3.0\%$ users of BR and $3.6\%$ users of CR. For
$GSGQ_{kNN}$, the improvement on query running time is even
more higher for SaR and SaR* because of the in-memory
optimizations (see
Fig.~\ref{fig:subfig:twknnelatency}). That is, compared to BR (resp. CR), SaR
and SaR* save $92.5\%$ (resp. $90.4\%$) and $93.5\%$ (resp.
$91.7\%$) of query running time. This indicates that by optimizing
the accessing order of the entries based on the CBRs, a greater
performance improvement can be achieved.

\begin{figure}
  \centering
  \subfigure[$GSGQ_{rkNN}$ (k=10, c=2)]{
    \label{fig:subfig:twknlatency} 
    \includegraphics[width=0.48\linewidth]{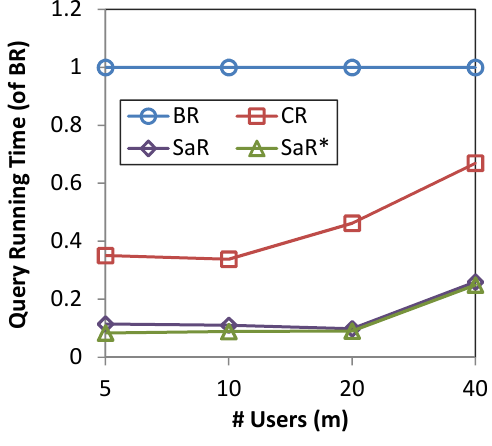}}
  \subfigure[$GSGQ_{rkNN}$ (k=10, c=2)]{
    \label{fig:subfig:twknlatency} 
    \includegraphics[width=0.48\linewidth]{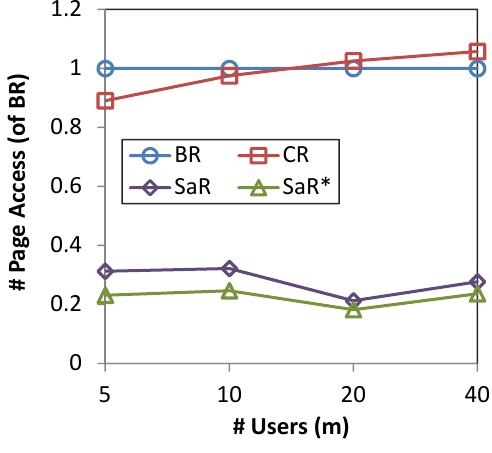}}
  \caption{Overall performance comparison on Twitter-2010 with different user \#.}
  \label{fig:knimpact} 
\end{figure}

Finally, comparing Fig.~\ref{fig:twoverallperform} to
Fig.~\ref{fig:overallperform} and Fig.~\ref{fig:dpoverallperform},
we can see that our methods gain a higher improvement over CR on
Twitter-2010 than on Gowalla and Dianping. This is because
Twitter-2010 has a denser social network and more diverse locations,
thus limiting the pruning power of the core numbers and making it
harder to process a GSGQ. As a further investigation on the impact of the social
graph with different sizes and density, we choose subsets of users in Twitter-2010 from 5M to 40M, and Table~\ref{tab:density} shows the average degrees and core numbers of these induced subgraphs. Fig.~\ref{fig:knimpact} plots the performance comparison of $GSGQ_{rkNN}$ queries on these social graphs. We can see that as the graph density grows, the performance gap
between CR and SaR/SaR* increases, because less pruning power can be
obtained from the core numbers. Compared to BR, SaR and SaR* retain
the pruning power and reduce the page access by roughly the same
ratio. The query running time of BR increases on the graph of
40M users because there is a jump of the graph density from 20M
users to 40M users and thus less time saving can be achieved in the
in-memory processing. To conclude, the pruning power of SaR and
SaR*, mainly contributed by the social relations in CBRs, benefits
more from larger and denser social networks.

\begin{table}
\centering \caption{Density of Twitter-2010 with different user \#.}
\label{tab:density}
\begin{tabular}{|l|c|c|c|c|} \hline
\hline User \# (m) & 5 & 10 & 20 & 40 \\
\hline Avg. degree & 1.957 & 2.613 & 4.783 & 28.556 \\
\hline Avg. core num. & 1.066 & 1.461 & 2.463 & 14.480 \\
\hline
\end{tabular}
\end{table}

\begin{figure}
  \centering
  \subfigure[Gowalla]{
    \label{fig:subfig:rangeclatency} 
    \includegraphics[width=0.48\linewidth]{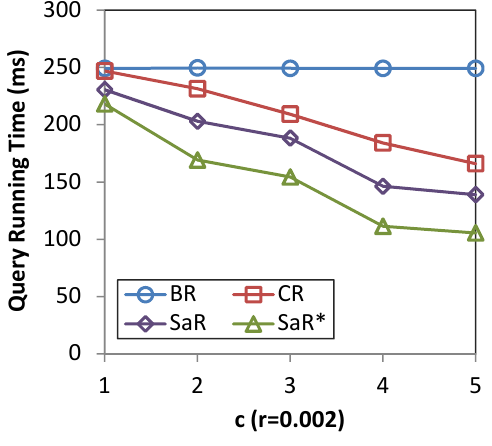}}
  \subfigure[Twitter-2010]{
    \label{fig:subfig:twrangeclatency} 
    \includegraphics[width=0.48\linewidth]{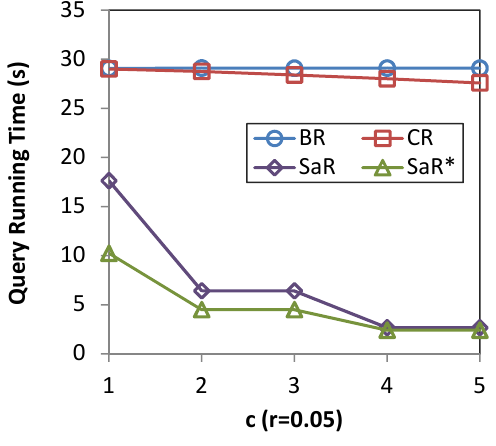}}
  \caption{Query running time of the methods for $GSGQ_{range}$ queries with different $c$ settings.}
  \label{fig:rcimpact} 
\end{figure}

\begin{figure}
  \centering
  \subfigure[Gowalla]{
    \label{fig:subfig:rangerlatency} 
    \includegraphics[width=0.48\linewidth]{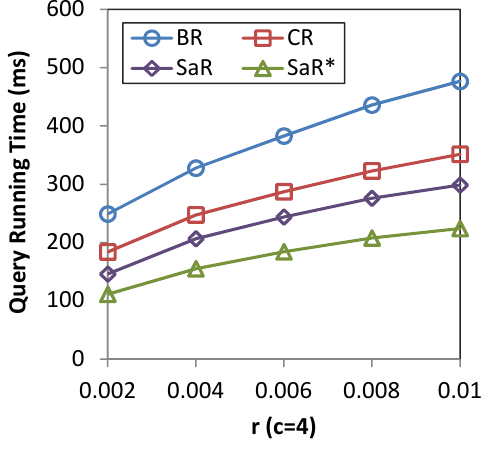}}
  \subfigure[Twitter-2010]{
    \label{fig:subfig:twrangerlatency} 
    \includegraphics[width=0.48\linewidth]{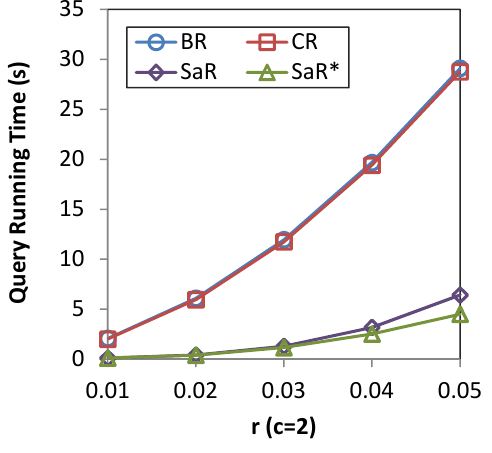}}
  \caption{Query running time of the methods for $GSGQ_{range}$ queries with different $r$ settings.}
  \label{fig:rrimpact} 
\end{figure}

\subsection{$GSGQ_{range}$ Processing}
For a $GSGQ_{range}$ $Q_{gs}=(v, r, c)$, Fig.~\ref{fig:rcimpact}
shows the performance with different $c$ settings on Gowalla
and Twitter-2010. All methods except BR incur shorter query running
time for a larger $c$. The performance gap between BR and the other methods increases as $c$ grows. This is because more users and
index nodes can be pruned in CR, SaR, and SaR* for a large $c$.
SaR and SaR* outperform CR in all cases. The improvement reduces a little at $c=3$ and $c=5$ because only
approximate CBRs (corresponding to $c=2$ and $c=4$, respectively) are used for query processing in these cases (recall that only the CBRs with respect to exponential
minimum degree constraints are stored).
Moreover, SaR* benefits more from the index than CR and SaR, as it
groups the users based on both spatial and social closenesses,
making the pruning of index nodes and user pages more powerful. As for various settings of query range $r$ (see
Fig.~\ref{fig:rrimpact}), the performance of all
methods degrades when $r$ grows, because more users within the range
need to be checked. In terms of query running time, SaR and SaR*
perform much better than the other two methods. Moreover,
SaR* has the best performance and thus is the
most favorable approach.



\begin{figure}
  \centering
  \subfigure[Gowalla]{
    \label{fig:subfig:knnclatency} 
    \includegraphics[width=0.48\linewidth]{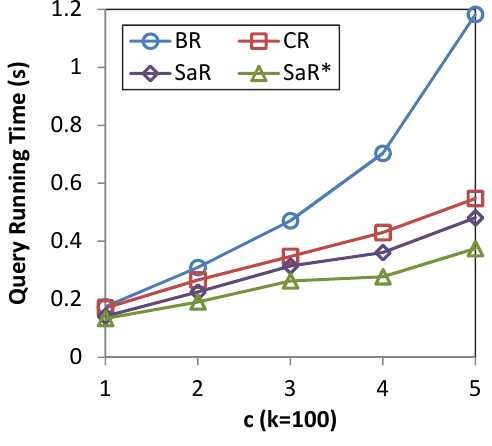}}
  \subfigure[Twitter-2010]{
    \label{fig:subfig:twknnclatency} 
    \includegraphics[width=0.48\linewidth]{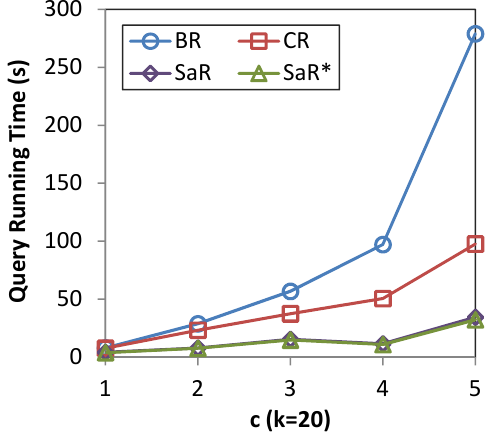}}
  \caption{Query running time of the methods for $GSGQ_{rkNN}$ queries
  with different $c$ settings.}
  \label{fig:kcimpact} 
\end{figure}

\begin{figure}
  \centering
  \subfigure[Gowalla]{
    \label{fig:subfig:knnklatency} 
    \includegraphics[width=0.48\linewidth]{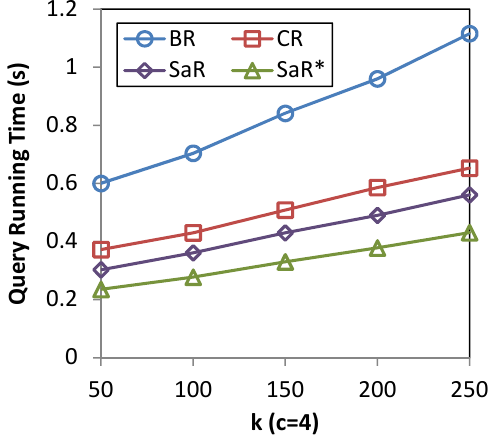}}
  \subfigure[Twitter-2010]{
    \label{fig:subfig:twknnklatency} 
    \includegraphics[width=0.48\linewidth]{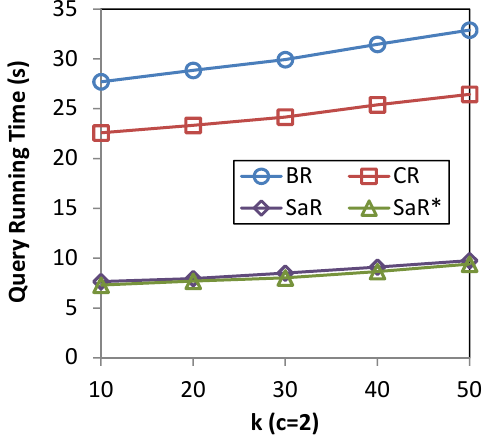}}
  \caption{Query running time of the methods for $GSGQ_{rkNN}$ queries
  with different $k$ settings.}
  \label{fig:kkimpact} 
\vspace*{0.1in}
\end{figure}

\subsection{$GSGQ_{rkNN}$ Processing}
This subsection investigates the performance of the methods for
$GSGQ_{rkNN}$ under various $c$ and $k$ settings. As we observed
similar performance trends for $GSGQ_{kNN}$ under these settings, we
omit the details on $GSGQ_{kNN}$ here.

For a $GSGQ_{rkNN}$ $Q_{gs}=(v, rkNN, c)$, Fig.~\ref{fig:kcimpact}
shows the performance with different $c$ settings on Gowalla and
Twitter-2010. All methods incur higher query running time for a
larger $c$. This is because a large $c$ tightens the social
constraint of $GSGQ_{rkNN}$ and thus more users need to be visited.
Similar to $GSGQ_{range}$, the performance gaps between SaR*
and the other two methods increase as $c$ grows. For a larger $c$,
the candidate users for $GSGQ_{rkNN}$
processing tend to share similar CBRs. Thus, the social-aware user
organization of SaR* can effectively reduce the page accesses.


Fig.~\ref{fig:kkimpact} shows the performance with different $k$
settings on Gowalla and Twitter-2010. Compared to $c$, the increment
of $k$ causes only a moderate increase in cost. SaR and SaR*
beat BR and CR for all $k$ settings and the performance gaps become larger
as $k$ grows. This implies that the pruning techniques of SaR and SaR* are scalable
to large user groups.

\begin{table}
\centering \caption{Average \# of updated CBRs w.r.t. batch update
size.} \label{tab:update}
\begin{tabular}{|l|c|c|c|c|c|c|}
\hline Upd. Size ($k$) & 1 & 3 & 10 & 30 & 100 & 300 \\
\hline Gowalla & 13.14 & 5.71 & 2.27 & 1.05 & 0.45 & 0.16 \\
\hline Upd. Size ($k$) & 10 & 30 & 100 & 300 & 1000 & \\
\hline Twitter-2010 & 2927.7 & 1137.3 & 346.1 & 115.4 & 34.6 & \\
\hline
\end{tabular} 
\end{table}


\subsection{Update Performance of SaR-trees}
\begin{figure}
  \centering
  \subfigure[Query Running Time]{
    \label{fig:subfig:updqt} 
    \includegraphics[width=0.48\linewidth]{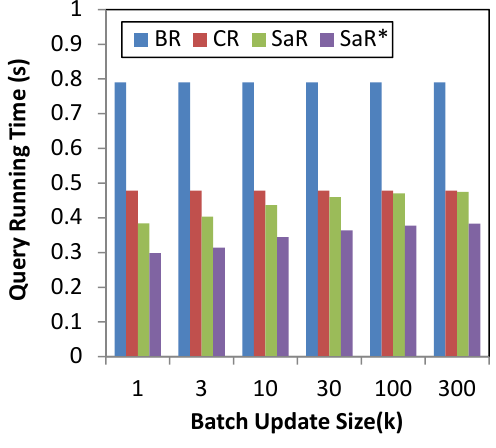}}
  \subfigure[Page Access Cost]{
    \label{fig:subfig:updqn} 
    \includegraphics[width=0.48\linewidth]{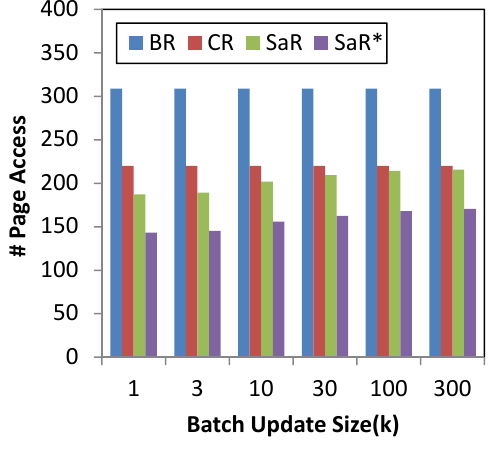}}
  \caption{The performance of the lazy update model on Gowalla.}
  \label{fig:update} 
\end{figure}

\begin{figure}
  \centering
  \subfigure[Query Running Time]{
    \label{fig:subfig:twupdqt} 
    \includegraphics[width=0.48\linewidth]{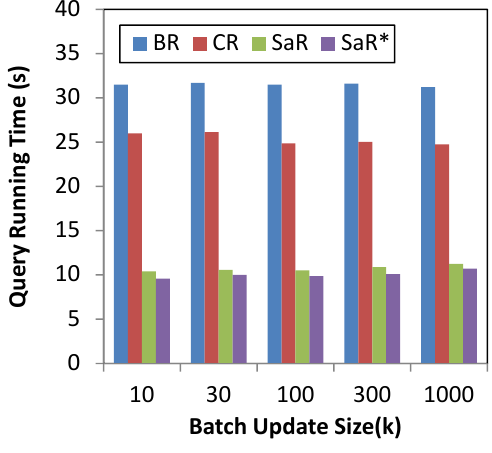}}
  \subfigure[Page Access Cost]{
    \label{fig:subfig:twupdqn} 
    \includegraphics[width=0.48\linewidth]{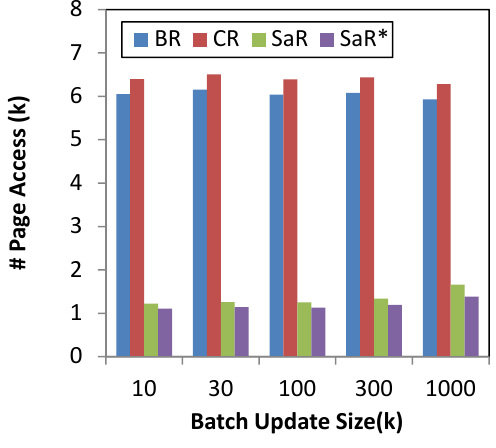}}
  \caption{The performance of the lazy update model on Twitter-2010.}
  \label{fig:twupdate} 
\end{figure}

This section investigates the update performance of SaR-trees. We
take the locations of user check-ins along the timeline of Gowalla
and Twitter-2010 to generate location updates (where the new
check-ins for Twitter-2010 are randomly generated with the maximum distance
0.0015 from the last ones) and randomly insert new edges
to generate social updates on users. Due to the fact that social
updates are relatively infrequent in real social networks
\cite{lesk}, the proportion of social updates is set to $5\%$. We
first investigate the effect of batch update size. In general, the
average amortized update time decreases as more updates are applied
in batch processing. This is mainly because fewer CBRs, on
average, are required to update as summarized in
Table~\ref{tab:update}. Fig.~\ref{fig:update} (resp. Fig.~\ref{fig:twupdate}) shows the
performance for the $GSGQ_{rkNN}$ queries with default settings
under different batch update sizes on Gowalla (resp. Twitter-2010).
We can see that the performance of SaR and SaR* degrades as the
batch update size grows, which is mainly because more CBRs are
invalidated by the updates of $M$ and less pruning power could be
achieved (yet still better than BR or CR).


\begin{figure}
  \centering
  \subfigure[Batch Update Size = $30k$]{
    \label{fig:subfig:thrat} 
    \includegraphics[width=0.48\linewidth]{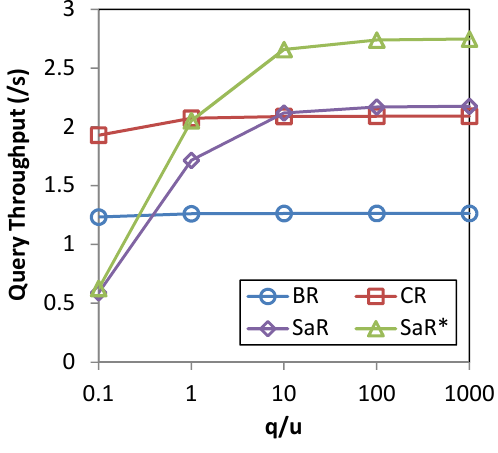}}
  \subfigure[$q/u = 10$]{
    \label{fig:subfig:thupd} 
    \includegraphics[width=0.48\linewidth]{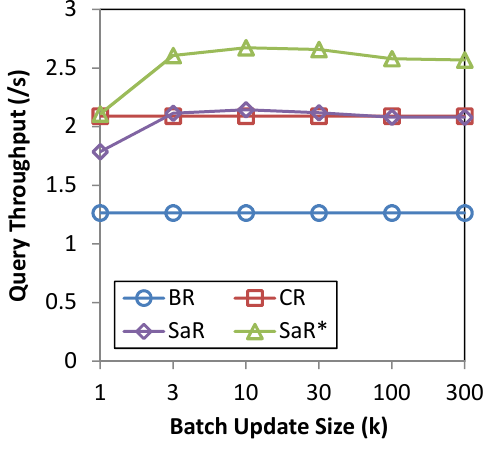}}
  \caption{The query throughput of the methods on Gowalla.}
  \label{fig:throughput} 
\end{figure}

\begin{figure}
  \centering
  \subfigure[Batch Update Size = $300k$]{
    \label{fig:subfig:twthrat} 
    \includegraphics[width=0.48\linewidth]{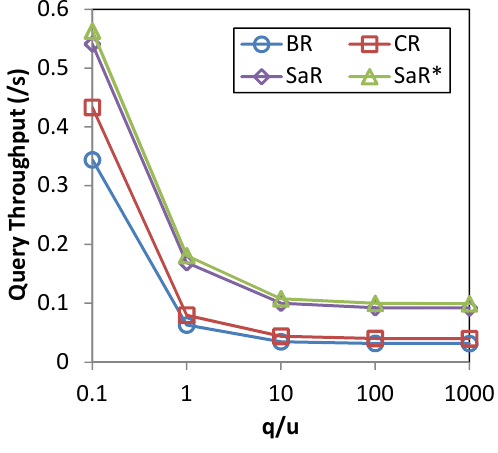}}
  \subfigure[$q/u = 10$]{
    \label{fig:subfig:twthupd} 
    \includegraphics[width=0.48\linewidth]{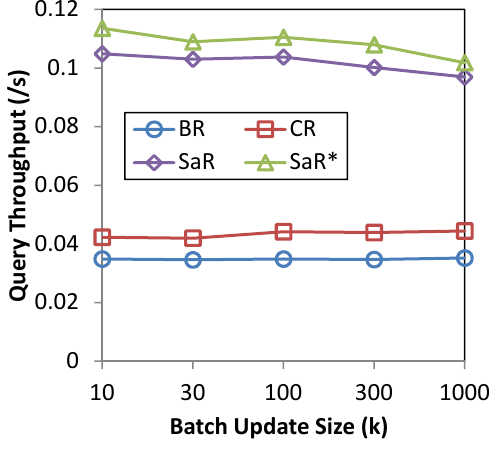}}
  \caption{The query throughput of the methods on Twitter-2010.}
  \label{fig:twthroughput} 
\end{figure}

To further measure the impact of updates on query processing, we
generate workloads of mixed update and query requests (i.e., the
$GSGQ_{rkNN}$ queries with default settings).
Fig.~\ref{fig:subfig:thrat} shows the throughputs
under various query/update ratios (workloads) on Gowalla.
SaR* and SaR achieve higher throughputs than CR when the workload has
fewer updates, i.e., $q/u>1$ and $10$, respectively, because the
performance gain from query processing can compensate for the
additional CBR update cost.
Fig.~\ref{fig:subfig:thupd} shows the thoughputs under different batch
update sizes on Gowalla. We can see that SaR outperforms CR only for
a range of the batch update size. It is because large batch update size
leads to obvious performance degradation of SaR for GSGQ processing,
making it incapable to compensate for the CBR update cost any more.
In comparison, SaR* always achieves the highest throughput. This
can also be observed on Twitter-2010, as shown in Fig.~\ref{fig:twthroughput}.

\subsection{Case Study: SSGQ vs. GSGQ}
\begin{figure}
\centering
\includegraphics[width=\linewidth]{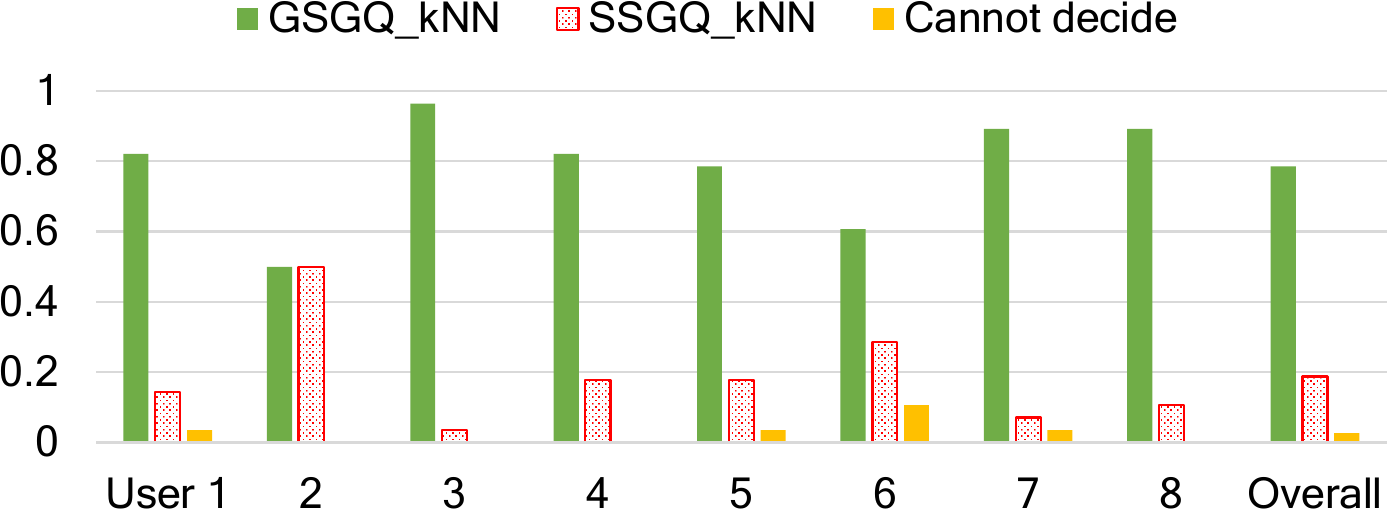}
\vspace{-0.1in}
\caption{\small Percentage of Participants' Choice for Each User}
\label{case_study}
\end{figure}

We also conducted a case study on the usefulness of GSGQ against SSGQ~\cite{yang12}. We randomly chose $8$ users from the Gowalla dataset and generated SSGQ and GSGQ kNN results under the following $4$ parameter settings (i.e., $2$ users under each setting): (1) $k=5$, $c=p=1$; (2) $k=5$, $c=p=2$; (3) $k=10$, $c=p=1$; (4) $k=10$, $c=p=2$ . For each user, the SSGQ result is visualized side by side with the GSGQ result in the context of Google Map and social relation of users. 28 participants were invited to give (blind) opinions on which result each user should choose for a group activity. Fig.~\ref{case_study} shows the comparison result. Of all $8$ users except for \#2 user, GSGQ is consistently chosen more often than SSGQ queries, and overall in $78$\% cases a participant chooses GSGQ results and in only $18$\% cases a participant chooses SSGQ results. This case study justifies our motivation of GSGQ as a more useful geo-social group query.

\section{Conclusion} \label{conclude}
This paper has studied geo-social group
queries (GSGQs) with minimum acquaintance constraints for large social networking services. Our main contribution is the design of two
social-aware index structures, namely SaR-tree and SaR*-tree. Based on them, we have developed efficient
algorithms to process various GSGQs, together with a number of optimization techniques. Extensive experiments on
real-world datasets demonstrate that our proposed methods
substantially outperform the baseline methods based on R-tree under
various system settings, and that such GSGQ services are feasible on a commodity server for large user populations. As for future work, we plan to extend GSGQs
to incorporate more sophisticated spatial queries such as skyline
and distance-based joins.

\section*{Acknowledgements}
This work was supported by National Natural Science Foundation of China (Grant No: 61572413 and U1636205), and Research Grants Council, Hong Kong SAR, China, under projects 12244916, 12201615, 12202414, 12200914, 15238116, and C1008-16G.




\end{document}